\documentclass[aps,prd,twocolumn,floatfix,nofootinbib,showpacs,superscriptaddress]{revtex4}

\usepackage{amsmath,amsfonts,amssymb,bm}
\usepackage{dsfont}
\usepackage{graphicx}
\usepackage{color}
\usepackage{soul}
\usepackage{slashed}

\usepackage[mathscr,scaled=1.15]{urwchancal}
\DeclareFontFamily{OT1}{pzc}{}
\DeclareFontShape{OT1}{pzc}{m}{it}%
{<-> s * [1.15] pzcmi7t}{}
\DeclareMathAlphabet{\mathpzc}{OT1}{pzc}{m}{it}

\definecolor{purple}{rgb}{0.5,0,0.5}
\definecolor{blue}{rgb}{0.0,0,0.9}

\begin{document}

\title{Contact-interaction Faddeev equation and, \emph{inter alia}, proton tensor charges}

\author{Shu-Sheng Xu}
\affiliation{Department of Physics, Nanjing University, Nanjing 210093, China}

\author{Chen Chen}
\affiliation{Hefei National Laboratory for Physical Sciences at the Microscale,
University of Science and Technology of China, Hefei, Anhui 230026, P. R. China}
\affiliation{Institute for Theoretical Physics and Department of Modern Physics,\\
University of Science and Technology of China, Hefei, Anhui 230026, P. R. China}

\author{Ian\;C.~Clo\"et}
\affiliation{Physics Division, Argonne National Laboratory, Argonne, Illinois 60439, USA}

\author{Craig D. Roberts}
\email[Corresponding author: ]{cdroberts@anl.gov}
\affiliation{Physics Division, Argonne National Laboratory, Argonne, Illinois 60439, USA}

\author{Jorge Segovia}
\affiliation{
Instituto Universitario de F\'{\i}sica Fundamental y Matem\'aticas (IUFFyM) Universidad de Salamanca, E-37008 Salamanca, Spain}

\author{Hong-Shi Zong}
\email[Corresponding author: ]{zonghs@nju.edu.cn}
\affiliation{Department of Physics, Nanjing University, Nanjing 210093, China}

\date{18 November 2015}

\begin{abstract}
A confining, symmetry-preserving, Dyson-Schwinger equation treatment of a vector$\,\otimes\,$vector contact interaction is used to formulate Faddeev equations for the nucleon and $\Delta$-baryon in which the kernel involves dynamical dressed-quark exchange and whose solutions therefore provide momentum-dependent Faddeev amplitudes. These solutions are compared with those obtained in the static approximation and with a QCD-kindred formulation of the Faddeev kernel.  They are also used to compute a range of nucleon properties, amongst them: the proton's $\sigma$-term; the large Bjorken-$x$ values of separate ratios of unpolarised and longitudinally-polarised valence $u$- and $d$-quark parton distribution functions; and the proton's tensor charges, which enable one to directly determine the effect of dressed-quark electric dipole moments (EDMs) on neutron and proton EDMs.
\end{abstract}


\maketitle

\section{Introduction}
A Poincar\'e-covariant Faddeev equation for baryons was introduced in Ref.\,\cite{Cahill:1988dx}.  In principle, it sums all possible quantum field theoretical exchanges and interactions that can take place between the three dressed-quarks that characterise a baryon's valence-quark content; and first, rudimentary computations were described in Ref.\,\cite{Burden:1988dt}.  Numerous analyses of baryon properties using this framework have subsequently appeared, with the level of sophistication and breadth of application increasing steadily, \emph{e.g}.\ Refs.\,\cite{Buck:1992wz, Ishii:1995bu, Oettel:1998bk, Oettel:1999gc, Ishii:1998tw, Hecht:2002ej, Bentz:2007zs, Eichmann:2008ef, Cloet:2008re, Eichmann:2009qa, Eichmann:2011vu, Chen:2012qr, Eichmann:2012mp, Cloet:2013gva, Roberts:2013mja, Sanchis-Alepuz:2014wea, Cloet:2014rja, Pitschmann:2014jxa, Segovia:2014aza, Segovia:2015hra, Segovia:2015ufa}.  The computational effort required to solve the Faddeev equation has naturally increased as the kernels have become more complex, so that one can consume significant resources in formulating and tackling even straightforward problems \cite{Eichmann:2009qa, Eichmann:2011vu, Eichmann:2012mp}.

\begin{figure}[b]
\centerline{%
\includegraphics[clip,width=0.45\textwidth]{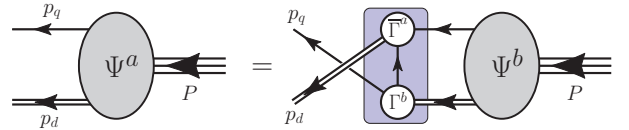}}
\caption{\label{figFaddeev} Poincar\'e covariant Faddeev equation.  $\Psi$ is the Faddeev amplitude for a baryon of total momentum $P= p_q + p_d$.  The shaded rectangle demarcates the kernel of the Faddeev equation: \emph{single line}, dressed-quark propagator; $\Gamma$,  diquark correlation amplitude; and \emph{double line}, diquark propagator.  (See Sec.\,\ref{SecFaddeev} for details.)}
\end{figure}

On the other hand, numerous practical and realistic simplifications present themselves.  Chief amongst these is suggested by the observation that bound-state studies which employ realistic quark-quark interactions \cite{Qin:2011dd, Binosi:2014aea} predict the appearance of nonpointlike colour-antitriplet quark$+$quark (diquark) correlations within baryons \cite{Cahill:1987qr, Maris:2002yu, Bender:2002as, Bhagwat:2004hn}.   Consequently, the baryon bound-state problem may be transformed into solving the linear, homogeneous matrix equation depicted in Fig.\,\ref{figFaddeev}.  The veracity of this approximation was established in Ref.\,\cite{Eichmann:2009qa}; and it has yielded a wide variety of novel predictions for baryon structure and interactions \cite{Chen:2012qr, Cloet:2013gva, Roberts:2013mja, Pitschmann:2014jxa, Segovia:2014aza, Segovia:2015hra, Segovia:2015ufa}.  Notably, empirical evidence supporting the presence of diquarks in the proton is accumulating \cite{Cates:2011pz, Cloet:2014rja, Segovia:2014aza,  Segovia:2015ufa}.

The diquarks within baryons are correlated by dressed-gluon exchange \cite{Segovia:2015ufa} and their structure is therefore described by a Bethe-Salpeter equation \cite{Cahill:1987qr, Maris:2002yu, Bender:2002as, Bhagwat:2004hn}.  These correlations are thus fundamentally different from the rudimentary, elementary diquarks introduced roughly fifty years ago in order to simplify treatment of the three-quark bound-state \cite{Lichtenberg:1967zz,Lichtenberg:1968zz}.  As highlighted by the shaded region in Fig.\,\ref{figFaddeev}, the two-body correlation predicted by modern Faddeev equation studies is not frozen: all dressed-quarks participate in all diquark clusters \cite{Segovia:2015ufa}.  The baryon spectrum produced by this Faddeev equation should therefore possess significant overlap with that of the three-quark constituent model and bear no simple relationship with that of the quark$+$elementary-diquark picture.

Given that solving the Faddeev equation can demand extensive numerical analysis, many authors have sought simplifications.  A drastic approach is to replace the Bethe-Salpeter kernel for the quark-quark system by a momentum-independent interaction, a motivation for which may be found in the Nambu--Jona-Lasinio model \cite{Nambu:1961tp, Nambu:2009zza}.  On the other hand, following a body of recent work \cite{GutierrezGuerrero:2010md, Roberts:2010rn, Roberts:2011wy, Roberts:2011cf, Wilson:2011aa, Chen:2012qr, Chen:2012txa, Pitschmann:2012by, Wang:2013wk, Segovia:2013rca, Segovia:2013uga} it may also be viewed differently.  Namely, since gluon propagation is characterised by a mass-scale $m_g \simeq 0.5\,$GeV, which screens infrared field modes \cite{Aguilar:2008xm, Brodsky:2008be, Aguilar:2009nf, Aguilar:2009pp, Binosi:2009qm, Boucaud:2011ug, Pennington:2011xs, Maas:2011se, Cucchieri:2011ig, Dudal:2012zx, Strauss:2012dg, Aguilar:2013xqa, Binosi:2014aea}, a vector$\,\otimes\,$vector contact-interaction can be a realistic representation of the quark-quark scattering kernel in QCD so long as the momentum scales being resolved are smaller than $m_g$.  Indeed, a symmetry-preserving Dyson-Schwinger equation (DSE) treatment of this interaction yields many results that are practically indistinguishable from those obtained with the most sophisticated kernels that have thus far been employed \cite{Roberts:2000aa, Maris:2003vk, Roberts:2007jh, Chang:2011vu, Bashir:2012fs, Cloet:2013jya}.

Recall now that quark exchange within the Faddeev equation kernel, highlighted by the shading in Fig.\,\ref{figFaddeev}, guarantees that no single dressed-quark is treated differently from the others.  This exchange also provides additional binding within the baryon.  Here, in the context of a contact vector$\,\otimes\,$vector quark-quark interaction, which produces diquark correlations amplitudes ($\Gamma$ in Fig.\,\ref{figFaddeev}) that are momentum-independent, dynamical quark-exchange is the only feature of the Faddeev kernel which can introduce momentum dependence into the baryon Faddeev amplitude, $\Psi$.  This complication, too, may be avoided if one employs a ``static approximation'' for the exchanged quark, \emph{viz}.\ \cite{Buck:1992wz}
\begin{equation}
\label{staticA}
S(p) = \frac{1}{i\gamma\cdot p +M } \to \frac{1}{M}\,,
\end{equation}
where $M \approx 0.4\,$GeV is the momentum-independent dressed-quark mass obtained in solving the gap equation using a contact interaction that provides an efficacious phenomenology.  Using Eq.\,\eqref{staticA}, or variants thereof, the Faddeev equation collapses to a simple algebraic expression, whose solution provides both a baryon's mass and a momentum-independent Faddeev amplitude, which expresses the relative strength of various possible diquark correlations in the bound-system \cite{Roberts:2011cf}.

Today, Eq.\,\eqref{staticA} is often a convenience, not a necessity.  It is employed by some authors because it can assist with the development of intuition about the bound three-valence-quark system and its properties.  Furthermore, where comparisons with more sophisticated studies can be made, such as for the nucleon, $\Delta$-baryon, and the Roper resonance \cite{Eichmann:2011vu, Segovia:2014aza, Segovia:2015hra}, the static-approximation provides an equally good description of baryon masses.  On the other hand, it provides a flawed description of hadron elastic and transition form factors because it leads to extreme hardening \cite{Segovia:2014aza}, and fails completely in describing the internal structure of the Roper resonance \cite{Segovia:2015hra}.

Herein we therefore analyse the nucleon and, to a lesser extent, $\Delta$-baryon Faddeev equations using the contact-interaction but eschewing the static-approximation, with an aim of determining whether the increased complexity brings benefits which outweigh the loss of simplicity.  We formulate the associated Faddeev equations in Sec.\,\ref{SecFaddeev} and describe a procedure for their solution in Sec.\,\ref{DefineFE}.  The results for the masses and amplitudes are presented and discussed in Sec.\,\ref{SecResults}.

As noted at the outset, the Faddeev equation is important because its solutions enable computation of numerous observable properties of baryons.  Hence, in Sec.\,\ref{SecCurrents} we present results for the nucleon $\sigma$-term, Bjorken-$x=1$ values for the separate ratios of unpolarised and longitudinally polarised $u$- and $d$-quark parton distribution functions (PDFs), the proton's tensor charges, and derived results for the neutron and proton electric dipole moments (EDMs).  We conclude in Sec.\,\ref{SecConclusion}.

\section{Contact-interaction and the Faddeev equation}
\label{SecFaddeev}
\subsection{Gap Equation}
\label{SecGap}
The basic pieces of the Faddeev equation in Fig.\,\ref{figFaddeev} are the dressed-quark and -diquark propagators, and the diquark Bethe-Salpeter amplitudes.  The dressed-quark propagator is obtained from a gap equation:
\begin{subequations}
\label{gendseN}
\begin{align}
S^{-1}(p) & =\,(i\gamma\cdot p + m^{\rm bm}) + \Sigma(p)\,,\\
\Sigma(p)& =  \int^\Lambda_{dq}\!\! g^2 D_{\mu\nu}(p-q)\frac{\lambda^a}{2}\gamma_\mu S(q) \frac{\lambda^a}{2}\Gamma_\nu(q,p) .
\end{align}
\end{subequations}
Here $D_{\mu\nu}$ is the gluon propagator; $\Gamma_\nu$, the quark-gluon vertex; $\int^\Lambda_{dq}$, a symbol representing a Poincar\'e invariant regularisation of the four-dimensional integral, with $\Lambda$ the regularisation mass-scale; and $m^{\rm bm}(\Lambda)$ is the current-quark bare mass.  We employ a confining, symmetry-preserving DSE treatment of a vector$\,\otimes\,$vector contact-interaction, which is implemented by writing
\begin{equation}
\label{njlgluon}
g^2 D_{\mu \nu}(p-q) = \delta_{\mu \nu} \frac{4 \pi \alpha_{\rm IR}}{m_G^2}\,,
\end{equation}
where $m_G=0.5\,$GeV is a gluon mass-scale typical of QCD and the fitted parameter $\alpha_{\rm IR} = 0.36\, \pi$ is commensurate with contemporary estimates of the zero-momentum value of a running-coupling in QCD \cite{Binosi:2014aea}.  This interaction is embedded in a rainbow-ladder (RL) truncation of the DSEs  \cite{Munczek:1994zz,Bender:1996bb}, \emph{viz}.\ one writes
\begin{equation}
\label{RLvertex}
\Gamma_{\nu}(p,q) =\gamma_{\nu}
\end{equation}
in the gap equation and in the subsequent construction of the Bethe-Salpeter kernels.  Notably, with realistic interactions \cite{Qin:2011dd}, the RL truncation is known to be reliable for properties of the following ground-states: isospin-nonzero-pseudoscalar- and vector-mesons, and octet and decuplet baryons \cite{Roberts:2000aa, Maris:2003vk, Roberts:2007jh, Chang:2011vu, Bashir:2012fs, Cloet:2013jya}.

Using Eqs.\,\eqref{njlgluon}, \eqref{RLvertex}, the gap equation becomes
\begin{equation}
 S^{-1}(p) =  i \gamma \cdot p + m +  \frac{16\pi}{3}\frac{\alpha_{\rm IR}}{m_G^2} \int^\Lambda_{dq}\!\,
\gamma_{\mu} \, S(q) \, \gamma_{\mu}\,.   \label{gap-1}
\end{equation}
We have assumed isospin symmetry, so that $m_u=m=m_d$ and strong interactions do not distinguish between $u$ and $d$-quarks, and use a Euclidean metric, which is detailed in Appendix\,A of Ref.\,\cite{Chen:2012qr}.  Equation~\eqref{gap-1} possesses a quadratic divergence; but if that is regularised in a Poincar\'e covariant manner, the solution is
\begin{equation}
\label{genS}
S(p)^{-1} = i \gamma\cdot p + M\,,
\end{equation}
where $M$ is momentum-independent and determined by
\begin{equation}
M = m + M\frac{4\alpha_{\rm IR}}{3\pi m_G^2} \int_0^\infty \!ds \, s\, \frac{1}{s+M^2}\,.
\end{equation}

We implement a confining regularisation of the contact-interaction by following Ref.\,\cite{Ebert:1996vx} and writing
\begin{align}
& \rule{-1em}{0ex} \frac{1}{s+M^2}  = \int_0^\infty d\tau\,{\rm e}^{-\tau (s+M^2)}
\label{ExplicitRS1}
\\
& \rule{-1em}{0ex} \rightarrow   \int_{\tau_{\rm uv}^2}^{\tau_{\rm ir}^2} d\tau\,{\rm e}^{-\tau (s+M^2)}
%
 = \frac{{\rm e}^{- (s+M^2)\tau_{\rm uv}^2}-{\rm e}^{-(s+M^2) \tau_{\rm ir}^2}}{s+M^2} \,, \label{ExplicitRS}
\end{align}
where $\tau_{\rm ir,uv}$ are, respectively, infrared and ultraviolet regulators.  This will be our definition of $\int^\Lambda_{dq}$.  Evidently, a finite value of $\tau_{\rm ir}=:1/\Lambda_{\rm ir}$ in Eq.\,(\ref{ExplicitRS}) implements confinement by ensuring the absence of quark production thresholds in colour singlet amplitudes.  It is worth adding to this remark.  Contemporary theory predicts that both quarks and gluons acquire mass distributions, which are large at infrared momenta \cite{Aguilar:2008xm, Brodsky:2008be, Aguilar:2009nf, Aguilar:2009pp, Binosi:2009qm, Boucaud:2011ug, Pennington:2011xs, Maas:2011se, Cucchieri:2011ig, Dudal:2012zx, Strauss:2012dg, Aguilar:2013xqa, Binosi:2014aea, Bhagwat:2003vw, Bowman:2005vx, Bhagwat:2006tu, Ayala:2012pb}.  The generation of these mass distributions leads to the emergence of a length-scale $\varsigma \approx 0.5\,$fm, whose existence is evident in all modern studies of dressed-gluon and -quark propagators and which signals a marked change in their analytic properties.  In this realisation, which has been canvassed by numerous authors (\emph{e.g}.\ Refs.\,\cite{Gribov:1999ui, Munczek:1983dx, Stingl:1983pt, Cahill:1988zi, Krein:1990sf, Aiso:1997au} and citations thereof), confinement is a dynamical process that may be realised via Eq.\,\eqref{ExplicitRS}.

\begin{table}[t]
\caption{\label{Tab:DressedQuarks}
Computed dressed-quark properties, required as input for the Bethe-Salpeter and Faddeev equations, and derived scalar and axial-vector diquark properties.  All results obtained with $\alpha_{\rm IR} =0.36 \pi$ and (in GeV) $\Lambda_{\rm ir} = 0.24\,$, $\Lambda_{\rm uv}=0.905$.
\emph{N.B}.\ These parameters are taken from the spectrum calculation of Ref.\,\protect\cite{Roberts:2011cf}.  
(Dimensioned quantities are listed in GeV.)}
\begin{center}
\begin{tabular*}
{\hsize}
{
c@{\extracolsep{0ptplus1fil}}
c@{\extracolsep{0ptplus1fil}}
c@{\extracolsep{0ptplus1fil}}
c@{\extracolsep{0ptplus1fil}}
c@{\extracolsep{0ptplus1fil}}
c@{\extracolsep{0ptplus1fil}}
c@{\extracolsep{0ptplus1fil}}
c@{\extracolsep{0ptplus1fil}}}\hline
$M_0$ & $m_u$  & $M_u$  & $m_{qq0^+}$ & $m_{qq1^+}$ & $E_{qq0}$ & $F_{qq0}$ & $E_{qq1}$\\\hline
0.36 & 0.007 & 0.37 & 0.78 & 1.06 & 2.74 & 0.31  & 1.30
\\\hline
\end{tabular*}
\end{center}
\end{table}

The interaction in Eq.\,(\ref{njlgluon}) does not define a renormalisable theory and hence $\Lambda_{\rm uv}:=1/\tau_{\rm uv}$ cannot be removed.  Instead, it plays a dynamical role, setting the scale of all dimensioned quantities.  Using Eq.\,\eqref{ExplicitRS}, the gap equation becomes
\begin{equation}
M = m+ M\frac{4\alpha_{\rm IR}}{3\pi m_G^2}\,\,{\cal C}^{\rm iu}(M^2)\,,
\label{gapactual}
\end{equation}
where ${\cal C}^{\rm iu}(\omega)/\omega = \overline{\cal C}^{\rm iu}(\omega) = \Gamma(-1,\omega \tau_{\rm uv}^2) - \Gamma(-1,\omega \tau_{\rm ir}^2)$, with $\Gamma(\alpha,y)$ being the incomplete gamma-function.  Solutions are listed in Table~\ref{Tab:DressedQuarks}.
\emph{N.B}.\ It is a feature of Eq.\,\eqref{gapactual} that in the chiral limit, $m=m_0=0$, a nonzero solution for $M_0:= \lim_{m_f\to 0} M_f$ is obtained so long as $\alpha_{\rm IR}$ exceeds a minimum value.  With $\Lambda_{\rm ir,uv}$ as specified in the Table, that value is $\alpha_{\rm IR}^c\approx 0.16\pi$.  This appearance of ``mass from nothing'' expresses the phenomenon of dynamical chiral symmetry breaking (DCSB), which is the source of more than 98\% of the mass of visible material in the Universe \cite{national2012NuclearS}.

\subsection{Diquark Correlations}
\label{SecDiquarks}
It has long been known \cite{Cahill:1987qr} that scalar and axial-vector diquark correlations are overwhelmingly dominant in the nucleon because all are positive-parity states and the correlation masses satisfy $m_{qq0^+}$, $m_{qq1^+}\lesssim m_N$.  Notably, the $\Delta$-baryon involves only axial-vector diquarks because it is impossible to build an isospin-$3/2$ state from a dressed-quark and an isospin-zero scalar diquark.

A contact interaction treatment of diquark correlations is detailed in Sec.\,2.2 of Ref.\,\cite{Chen:2012qr}.  In this case, the scalar and axial-vector diquarks are described by the following amplitudes:
\begin{subequations}
\begin{eqnarray}
\label{scqqbsa}
\Gamma^{0^+}_{qq}(P) &= &  i \gamma_5 E_{qq 0}(P) + \frac{1}{M} \gamma_5 \gamma\cdot P F_{qq 0}(P) \,,\rule{2em}{0ex}\\
i\Gamma_{qq\, \mu}^{1^+}(P) & = & i\gamma^T_\mu E_{qq 1}(P),\label{avqqbsa}
\end{eqnarray}
\end{subequations}
where $P_\mu \gamma^T_\mu = 0$.  These amplitudes do not depend on the quark-quark relative momentum; and in all applications they are canonically normalised ($q_+=q+P, S_{P_\mu}(q_+):=\partial S(q_+)/\partial P_\mu$):
\begin{subequations}
\begin{align}
P_\mu &= 2 {\rm tr}\!\! \int^\Lambda_{dq} \Gamma_{qq}^{0^+}(-P) S_{P_\mu}(q_+) \Gamma_{qq}^{0^+}(P) S(q)\,,\\
P_\mu &= \frac{2}{3} {\rm tr}\!\!\int^\Lambda_{dq} \Gamma_{qq\,\alpha}^{1^+}(-P)
S_{P_\mu}(q_+) \Gamma_{qq\,\alpha}^{1^+}(P) S(q)\,.
\label{avqqNorm}
\end{align}
\end{subequations}
The computed values of the correlation mass-scales and the constants which characterise the amplitudes are listed in Table~\ref{Tab:DressedQuarks}.

\subsection{Faddeev Equation}
\label{SecFaddeevEquation}
Numerous details relating to the Faddeev equation treatment of the nucleon and $\Delta$-baryon are provided elsewhere (\emph{e.g}.\ Sec.\,3 and Appendix~C in Ref.\,\cite{Chen:2012qr}). Here we merely recapitulate the main ideas, as they are relevant to our analysis.

The nucleon Faddeev amplitude is
\begin{equation}
\label{PsiN}
\Psi = \Psi_1 + \Psi_2 + \Psi_3  \,,
\end{equation}
where the subscript identifies the bystander quark and, \emph{e.g}.\ $\Psi_{1,2}$ are obtained from $\Psi_3$ by a cyclic permutation of all quark labels.  As remarked above, the nucleon is composed from a sum of scalar and axial-vector diquark correlations:
\begin{equation}
\label{amplitude1} \Psi_3(p_j,\alpha_j,\tau_j) = {\cal N}_{\;3}^{0^+} + {\cal N}_{\;3}^{1^+},
\end{equation}
with $(p_j,\alpha_j,\tau_j)$ representing the momentum, spin and flavour labels of the
quarks constituting the bound state, and $P=p_1+p_2+p_3$ is the system's total momentum.

The scalar diquark piece in Eq.\,(\ref{amplitude1}) is
\begin{equation}
\label{scalarN}
\mathcal{N}_3^{0^+}(p_i,\alpha_i,\tau_i)=\left[ \Gamma^{0^+}\!(K) \right]^{\tau_1\tau_2}_{\alpha_1\alpha_2} \Delta^{0^+}(K) \left[ \mathcal{S}(l;P) u(P) \right]^{\tau_3}_{\alpha_3},
\end{equation}
where ${\cal S}$ is a $4\times 4$ Dirac matrix, which describes the relative quark--scalar-diquark momentum correlation within the nucleon, the spinor satisfies
\begin{equation}
\bar u(P)(i\gamma\cdot P+m_N)=0=(i\gamma\cdot P +m_N)u(P)\,,
\end{equation}
with $m_N$ the nucleon mass, $K=p_1+p_2=:p_{\{12\}}$, $l:=(-p_{\{12\}}+2p_3)$, and
\begin{equation}
\Delta^{0^+}(K)=\frac{1}{K^2+m^2_{qq{0^+}}}
\end{equation}
is the scalar diquark propagator.  The colour antisymmetry of $\Psi_3$ is implicit in $\Gamma^{J^P}\!\!$, with the Levi-Civita tensor, $\epsilon_{c_1 c_2 c_3}$, expressed via the antisymmetric Gell-Mann matrices, \emph{viz}. $\epsilon_{c_1 c_2 c_3}= (H^{c_3})_{c_1 c_2}$ when
\begin{equation}
\label{Hmatrices}
\{H^1=i\lambda^7,H^2=-i\lambda^5,H^3=i\lambda^2\}\,.
\end{equation}

The axial-vector component in Eq.\,\eqref{amplitude1} is
\begin{align}
\nonumber
&\mathcal{N}_3^{1^+}(p_i,\alpha_i,\tau_i) \\
&=\left[ t^i \Gamma_\mu^{1^+}(K) \right]^{\tau_1\tau_2}_{\alpha_1\alpha_2} \Delta_{\mu\nu}^{1^+}(K)\left[ \mathcal{A}_\nu^i(l;P)u(P) \right]^{\tau_3}_{\alpha_3},
\label{pseudovectorN}
\end{align}
where ${\cal A}_\nu$ is a $4\times 4$ Dirac matrix, which describes the relative quark--pseudovector-diquark momentum correlation within the nucleon, the symmetric isospin-triplet matrices are
\begin{equation}
t^+=\frac{1}{\sqrt{2}}(\tau^0+\tau^3), t^0=\tau^1, t^-=\frac{1}{\sqrt{2}}(\tau^0-\tau^3)
\end{equation}
and, with $T^1_{\mu\nu}(K) = \delta_{\mu\nu}+K_\mu K_\nu/m_{qq_{1^+}}$,
\begin{equation}
\Delta_{\mu\nu}^{1^+}(K)=\frac{1}{K^2+m_{qq{1^+}}^2} \, T^1_{\mu\nu}(K)\,.
\end{equation}

The $\Delta$-baryon contains only an axial-vector diquark:
\begin{equation}
\Psi_3^\Delta(p_i,\alpha_i,\tau_i)=\mathcal{D}_3^{1^+} ;
\end{equation}
and when computing the mass and Faddeev amplitude one can focus on the $\Delta^{++}$, owing to isospin symmetry, in which case:
\begin{equation}
\mathcal{D}_3^{1^+}=\left[ t^+ \Gamma_\mu^{1^+}(K) \right]^{\tau_1\tau_2}_{\alpha_1\alpha_2} \Delta_{\mu\nu}^{1^+}(K) \left[ \mathcal{D}_{\nu\rho}(l;P) u_\rho(P)\right]^{\tau_3}_{\alpha_3},
\end{equation}
where $\mathcal{D}_{\nu\rho}$ is a $4\times 4$ Dirac matrix, which describes the relative quark--pseudovector-diquark momentum correlation within the $\Delta^{++}$, and $u_\rho(P)$ is a Rarita-Schwinger spinor (see Appendix~A of Ref.\,\cite{Chen:2012qr} for details).

The Dirac-matrix structure of Eqs.\,\eqref{scalarN}, \eqref{pseudovectorN} may be expressed as follows:
\begin{subequations}
\label{SAStructure}
\begin{align}
\label{SStructure}
\mathcal{S}(l;P)&=\sum_{n=1}^2 s_n(l;P) \tau^n(l;P)\,,
\\
\mathcal{A}^i_\mu(l;P)&=\sum_{n=3}^8 a^i_n(l;P) \tau_\mu^n(l;P)\,,\; i =+,0\,,
\end{align}
\end{subequations}
where $s_n(l;P)$ and $a^i_n(l;P)$ are scalar functions of $l^2$, $l\cdot P$, and the $4\times 4$ matrices $\{\tau^i, \tau_\mu^j\,|\,i=1,2;j=3,\ldots,8\}$ are defined in Eq.\,\eqref{tauN} of Appendix\,\ref{AppFormulae}.  The $\Delta$-baryon amplitudes may be expressed similarly:
\begin{equation}
\label{Dstructure}
\mathcal{D}_{\nu\rho}(l;P)=\sum_{n=1}^8 f_n(l;P) \tau_{\nu\rho}^n(l;P)\,,
\end{equation}
where $f_n(l;P)$  are scalar functions and the $4\times 4$ matrices $\{\tau_{\nu\rho}^i \, | \, i=1,\ldots,8\}$ are defined in Eq.\,\eqref{tauD}.

One can now express the Faddeev equation for $\Psi_3$:
\begin{eqnarray}
\nonumber
\lefteqn{
 \left[ \begin{array}{r}
{\cal S}(k;P)\, u(P)\\
{\cal A}^i_\mu(k;P)\, u(P)
\end{array}\right]}\\
& =&  -\,4\,\int^\Lambda_{dl}\,{\cal M}(k,l;P)
\left[
\begin{array}{r}
{\cal S}(l;P)\, u(P)\\
{\cal A}^j_\nu(l;P)\, u(P)
\end{array}\right] ,\rule{1em}{0ex}
\label{FEone}
\end{eqnarray}
where
\begin{equation}
\label{calM} {\cal M}(k,l;P) = \left[\begin{array}{cc}
{\cal M}_{00} & ({\cal M}_{01})^j_\nu \\
({\cal M}_{10})^i_\mu & ({\cal M}_{11})^{ij}_{\mu\nu}\rule{0mm}{3ex}
\end{array}
\right] ,
\end{equation}
with
\begin{subequations}
\begin{eqnarray}
\nonumber
 {\cal M}_{00} &=& \Gamma^{0^+}\!(l_{qq})\,
S^{\rm T}(l_{qq}-k_q) \\
&& \times \,\bar\Gamma^{0^+}\!(-k_{qq})\,
S(l_q)\,\Delta^{0^+}(l_{qq}) \,, \rule{2em}{0ex}\\
\nonumber
({\cal M}_{01})^j_\nu &=& {\tt t}^j \,
\Gamma_\mu^{1^+}\!(l_{qq}) S^{\rm T}(l_{qq}-k_q)\,\\
&& \times \bar\Gamma^{0^+}\!(-k_{qq})\,
S(l_q)\,\Delta^{1^+}_{\mu\nu}(l_{qq}) , \rule{2.2em}{0ex} \label{calM01} \\
\nonumber
({\cal M}_{10})^i_\mu &=& \Gamma^{0^+}\!(l_{qq})\,
S^{\rm T}(l_{qq}-k_q)\,{\tt t}^i\, \\
&&\times \bar\Gamma_\mu^{1^+}\!(-k_{qq})\,
S(l_q)\,\Delta^{0^+}(l_{qq}) , \rule{2.2em}{0ex}\\
\nonumber
({\cal M}_{11})^{ij}_{\mu\nu} &=& {\tt t}^j\,
\Gamma_\rho^{1^+}\!(l_{qq})\, S^{\rm T}(l_{qq}-k_q)\,{\tt t}^i\,\\
&& \times  \bar\Gamma^{1^+}_\mu\!(-k_{qq})\,
S(l_q)\,\Delta^{1^+}_{\rho\nu}(l_{qq}) . \rule{2.2em}{0ex}\label{calM11}
\end{eqnarray}
\end{subequations}
where: $l_q=l+P/3$, $k_q=k+P/3$, $l_{qq}=-l+ 2P/3$,
$k_{qq}=-k+2P/3$; $\bar\Gamma=C^\dagger \, \Gamma(P)^{\rm T}\,C$, with $C=\gamma_2\gamma_4$ being the charge conjugation matrix, $C^\dagger=-C$; and the superscript ``T'' denotes a transposing of all matrix indices.

At this point one can use isospin symmetry to define $\mathcal{A}(k;P):= \mathcal{A}^0(k;P) = -\mathcal{A}^+(k;P)/\sqrt{2}$ and therewith simplify Eq.\,\eqref{FEone}:
\begin{align}
\nonumber
& \left[
\begin{array}{l}
\mathcal{S}(k;P)\, u(P)\\
\mathcal{A}_\mu(k;P) \,u(P)
\end{array}
\right]\\
& = -4 \int^\Lambda_{dl} \overline{\mathcal{M}}(k,l;P)
\left[
\begin{array}{l}
\mathcal{S}(l;P) \,u(P)\\
\mathcal{A}_\nu(l;P)\, u(P)
\end{array}
\right]\,,
\label{NuclFaddEq12}
\end{align}
where, with all entries referring to $i,j=0$,
\begin{eqnarray}
\overline{\mathcal{M}}(k,l;P)&=&
\left[
\begin{array}{ll}
\mathcal{M}_{00}		&3(\mathcal{M}_{01})_\nu	\\
(\mathcal{M}_{10})_\mu	&-(\mathcal{M}_{11})_{\mu\nu}
\end{array}
\right]\,.
\end{eqnarray}

The $\Delta$-baryon Faddeev equation is
\begin{align}
\nonumber
& \mathcal{D}_{\lambda\rho}(k;P)u_{\rho}(P)\\
& = 4 \int^\Lambda_{dl} \mathcal{M}^\Delta_{\lambda\mu}(k,l;P)\mathcal{D}_{\mu\sigma}(l;P)u_\sigma(P)\,,
\label{DeltaFaddEq1}
\end{align}
with
\begin{align}
\nonumber
 \mathcal{M}^\Delta_{\lambda\mu}&(k,l;P)= t^+\Gamma^{1^+}_\sigma(l_{qq})\\
&
\times S^T(l_{qq}-k_q)t^+ \bar\Gamma_\lambda(-k_{qq})S(l_q)\Delta^{1^+}_{\sigma\mu}(l_{qq})\,.
\end{align}

\section{Defining and Solving the Faddeev Equations}
\label{DefineFE}
\subsection{Scalar Equations}
Since we are not going to employ the so-called static-approximation, Eq.\,\eqref{staticA}, the Faddeev equations must be solved numerically, with momentum-dependent results for all the functions in Eqs.\,\eqref{SAStructure}, \eqref{Dstructure}.  To begin that process, we define a set of Dirac-matrix-valued projection operators, Eqs.\,\eqref{ProjectionN}, \eqref{ProjectionD} in Appendix~\ref{AppFormulae}, in order to convert the linear, homogeneous, Dirac-matrix-valued Faddeev equations into a set of equations for these scalar functions.  Left-multiplying Eq.\,\eqref{NuclFaddEq12} with the projection operators, right-multiplying with $\bar u(P)$, and forming the spinor trace, one obtains:
\begin{align}
\label{ScalarN}
\phi_m^N(k;P) = -4 \int^\Lambda_{dl} \sum_{n=1}^8 {\mathpzc K}_{\;N}^{mn}\,\phi_n^N(l;P)\,,
\end{align}
where each kernel entry ${\mathpzc K}_{\;N}^{mn}$ is a function of $k^2$, $l^2$, $k\cdot P$, $l\cdot P$, $k\cdot l$, and we have defined
\begin{align}
&\phi_m^N(k;P):=
\left\{
\begin{array}{ll}
s_m(k;P)\,, &m=1,2	\\
a_m(k;P)\,, &m=3,\ldots,8
\end{array}\right. \,,\\
&{\mathpzc K}_{\;N}^{mn}:=
\left\{
\begin{array}{ll}
{\mathpzc K}_{\;00}^{mn},		& m=1,2 \,,\; n=1,2	\\
{\mathpzc K}_{\;01}^{mn},	    &m=1,2 \,,\; n=3,\ldots,8 \\
{\mathpzc K}_{\;10}^{mn},	    &m=3,\ldots,8\,,\; n=1,2 \\
{\mathpzc K}_{\;11}^{mn},	    &m=3,\ldots,8\,,\; n=3,\cdots,8
\end{array}
\right. \,,
\end{align}
where the form of the entries is obvious, given the description above and Eq.\,\eqref{Lplus}, \emph{e.g}.\
\begin{align}
\nonumber {\mathpzc K}_{\;00}^{mn} =
{\rm tr}\,& \bar{{\tau}}_m(k;P)\Gamma^{0^+}(l_{qq})S^T(l_{qq}-k_q)\bar\Gamma^{0^+}(-k_{qq})\\
& \times S(l_q)\Delta^{0^+}(l_{qq}) \tau_n(l;P)\Lambda_+(P)\,. \label{K00N}
\end{align}

Following a similar procedure, an analogous equation for the $\Delta$-baryon is readily obtained:
\begin{equation}
\label{ScalarD}
\phi^\Delta_m(k;P) =
4\int^\Lambda_{dl} \sum_{n=1}^8 {\mathpzc K}_{\;\Delta}^{mn} \phi_n^\Delta(l;P)\,,
\end{equation}
where
\begin{align}
\nonumber {\mathcal K}_{\;\Delta}^{mn} ={\rm tr} \, & \bar{\tau}^m_{\lambda\nu}(k;P)\Gamma_\rho^{1^+}(l_{qq})S^T(l_{qq}-k_q) \bar\Gamma_\lambda^{1^+}(-k_{qq}) \\
& \times S(l_q)\Delta_{\rho\sigma}^{1^+}(l_{qq}) \tau^n_{\sigma\kappa}(l;P)\mathcal{R}^{\Delta}_{\kappa\nu}(P)\,.
\end{align}

\subsection{Regularised Equations}
\label{SecRegEq}
As is typical for bound-state equations founded on a contact interaction, Eqs.\,\eqref{ScalarN}, \eqref{ScalarD} involve divergences, which we tame by using the confining regularisation procedure described in connection with the gap equation, Eqs.\,\eqref{ExplicitRS1}, \eqref{ExplicitRS}.  For illustration, consider the first entry on the right-hand-side (rhs) in the equation for $\phi_1^N(k;P)$, \emph{viz}.\
\begin{equation}
\int^\Lambda_{dl} {\mathpzc K}_{\;N}^{11}\,\phi_1^N(l;P)
= \int^\Lambda_{dl} \frac{{\mathpzc N}^{11}(k,l;P)}{D_1 D_2 D_3}\phi_1^N(l;P)
\,,
\end{equation}
where, using Eq.\,\eqref{K00N},
\begin{subequations}
\begin{align}
\nonumber
{\mathpzc N}^{11} & =
{\rm tr}\, \bar{{\tau}}_1(k)\Gamma^{0^+}(l_{qq})[-i(\gamma\cdot{l}_{qq} -\gamma\cdot{k}_q)+M]^{\rm T} \\
& \times \bar\Gamma^{0^+}(-k_{qq})(-i\gamma\cdot{l}_q+M) \tau_1(l;P)\Lambda_+(P) \,,\\
D_1 & =  l_q^2+M^2\,,\\
D_2 & =  (l_{qq}-k_q)^2+M^2\,,\\
D_3 & = l_{qq}^2+m_{qq0^+}^2\,.
\end{align}
\end{subequations}

The explicit form for ${\mathpzc N}^{11}$ is cumbersome, so we do not include it here; but it is worth detailing our treatment of the denominator product:
\begin{equation}
\frac{1}{D_1 D_2 D_3} = 2\int_0^1 dx \, dy \, x \frac{1}{D^3}\,,
\end{equation}
where
\begin{subequations}
\begin{align}
\nonumber
D & =(1-x) D_1 + x(1-y)D_2+x y D_3\\
& = (l+xyk-(x-\eta(1+xy))P)^2+\omega \,, \\
\nonumber
\omega &= M^2[1-x(1-y)] + xy [ k^2-2k\cdot P (1-\eta-x) \\
\nonumber
& \quad - (2\eta k\cdot P + k^2)xy]  + x[ m^2_{qq0^+}(1-y) \\
& \quad - m_N^2 (1-[2-\eta]\eta y+x[1-\eta y]^2) ]\,.
\end{align}
\end{subequations}

Assuming a Poincar\'e-covariant regularisation procedure, one may shift the integration variable as follows $l_\mu\rightarrow l_\mu-xy k_\mu +(x-\eta(1+xy))P_\mu$, so that
\begin{align}
\nonumber
{\mathpzc N}^{11}&(k,l;P)  \rightarrow \tilde{\mathpzc N}^{11}(k,l;P)\\
& ={\mathpzc N}^{11}(k,l-xy k +(x-\eta(1+xy))P;P)
\end{align}
and hence
\begin{align}
\nonumber
&\int^\Lambda_{dl} \frac{{\mathpzc N}^{11}(k,l;P)}{D_1 D_2 D_3}\,\phi_1^N(k;P)\\
& = 2\int_0^1 dx \, dy \, x
\int^\Lambda_{dl} \frac{\tilde{\mathpzc N}^{11}(k,l;P)}{(l^2+\omega)^3}\phi_1(l;P)\,.
\label{FE2}
\end{align}
\emph{N.B}.\ After the change-of-variables, $l\cdot P = 0$; and in the computation of $\tilde{\mathpzc N}^{11}(k,l;P)$ one may then follow Ref.\,\cite{Roberts:2011cf} and set:
\begin{equation}
k_{qq}^2 \rightarrow -m_{qq}^2,\,
l_{qq}^2 \rightarrow -m_{qq}^2,\,
k_{qq\,\mu} \rightarrow \tfrac{2}{3}P_\mu,\,
l_{qq\,\mu} \rightarrow \tfrac{2}{3}P_\mu.
\end{equation}

At this point we use Eqs.\,\eqref{ExplicitRS1}, \eqref{ExplicitRS} and infer
\begin{align}
\frac{1}{{\mathpzc z}^{n+1}} \rightarrow {\mathpzc E}_n^{\rm iu}({\mathpzc z})
 := \frac{(-1)^n}{n!} \frac{d^n}{d\omega^n}
\frac{e^{-\tau_{uv}^2{\mathpzc z}}-e^{-\tau^2_{ir}{\mathpzc z}}}{{\mathpzc z}} \,,
\end{align}
in which case Eq.\,\eqref{FE2} becomes
\begin{align}
\nonumber
\int^\Lambda_{dl} &\frac{{\mathpzc N}^{11}(k,l;P)}{D_1 D_2 D_3}\,\phi_1^N(k;P)
= 2\int_0^1 dx \, dy \, x\\
&\times
\int_{dl}^\infty
{\mathpzc N}^{11}(k,l;P) \, {\mathpzc E}_2^{\rm iu}(l^2+\omega)\phi_1^N(k;P)\,,
\end{align}
where $\int_{dl}^\infty=\int d^4 l/(2\pi)^4$ represents the unbounded four-dimensional momentum integral.  The integral on the rhs is now free from ultraviolet divergences and exhibits no quark production thresholds, \emph{i.e}.\ confinement is realised.

\subsection{Solving the Regularised Equations}
Full implementation of the procedures illustrated above yields a collection of well-defined linear, homogeneous integral equations, one set for the nucleon Faddeev amplitude and another set for the $\Delta$-baryon.  These equations only have solutions at discrete, separated values of $P^2$.  As usual, therefore, we consider a modified equation, which takes the form, using the nucleon as an example:
\begin{align}
\label{ScalarNL}
\lambda(P^2) \phi_m^N(k;P) = -4 \int^\Lambda_{dl} \sum_{n=1}^8 {\mathpzc K}_{\;N}^{mn}\,\phi_n^N(l;P)\,.
\end{align}
Equation\,\eqref{ScalarNL} has at least one solution for every value of $P^2$; and Faddeev equation solutions for physical bound-states are obtained at those values of $P^2$ for which $\lambda(P^2)=1$.  Any sensible numerical procedure may now be used to locate  bound-state masses and determine the associated Faddeev amplitudes.

\begin{table}[t]
\caption{\label{TabFEMasses}
Computed masses of the nucleon and $\Delta$-baryon, obtained with the contact interaction parameters listed in Table~\ref{Tab:DressedQuarks}.  The quantities $g_{N,\Delta}$ are described in connection with Eq.\,\eqref{staticAR}.
(Dimensioned quantities are listed in GeV.)}
\begin{center}
\begin{tabular*}
{\hsize}
{
l@{\extracolsep{0ptplus1fil}}
l@{\extracolsep{0ptplus1fil}}
c@{\extracolsep{0ptplus1fil}}
c@{\extracolsep{0ptplus1fil}}}\hline
Herein &       & $g_N=1$, $g_\Delta=1$ & $g_N=1.28$, $g_\Delta=1.73$ \\\hline
       & $m_N$ & 1.30 & 1.14 \\
       & $m_\Delta$ & 1.65 & 1.39
       \\\hline
Ref.\,\cite{Chen:2012qr}
       &       & $g_N=1$, $g_\Delta=1$ & $g_N=1.18$, $g_\Delta=1.56$ \\\hline
       & $m_N$ & 1.27 & 1.14 \\
       & $m_\Delta$ & 1.60 & 1.39
       \\\hline
\end{tabular*}
\end{center}
\end{table}

\section{Results from the Faddeev Equations}
\label{SecResults}
We solved the Faddeev equations using the contact-interaction parameters described in connection with Table~\ref{Tab:DressedQuarks}.  The computed masses are listed in Table~\ref{TabFEMasses}.  The first column of numerical entries compares our results, obtained with a full treatment of the contact-interaction, with those produced using the original static-approximation, Eq.\,\eqref{staticA}.  It provides a surprise, \emph{viz}.\ any additional attraction introduced by our complete treatment of the quark-exchange kernel is compensated by the appearance of additional spin-orbit repulsion, expressed in the momentum-dependent Faddeev amplitude by the presence of the components $s_2(k;P)$, $a_{4,6,7,8}(k;P)$ in Eqs.\,\eqref{SAStructure}.  The calculated masses are therefore almost unaffected by eliminating the static-approximation.

The numerical values of the calculated masses are shifted roughly 25\% above their respective empirical values.  It is appropriate that the Faddeev equation in Fig.\,\ref{figFaddeev} should yield masses that are larger than experiment because, as explained elsewhere \cite{Eichmann:2008ae, Eichmann:2008ef}, the kernel in Fig.\,\ref{figFaddeev} omits all those resonant contributions which may be associated with the meson-baryon final-state interactions that are resummed in dynamical coupled channels models \cite{Suzuki:2009nj,Kamano:2013iva,Doring:2014qaa} in order to transform a bare-baryon into the observed state.  Our Faddeev equation should therefore be understood as producing the dressed-quark core of the bound-state, not the completely-dressed and hence observable object.  The problem here is that the shift is too large.

Analysis of the effect of meson-baryon final-state interactions indicates that they typically produce a 15\% reduction in nucleon and $\Delta$-baryon quark-core masses.  As noted in Ref.\,\cite{Roberts:2011cf}, a Faddeev equation kernel capable of producing more realistic quark-core masses can be obtained through a modest modification of the quark exchange kernel, which is herein implemented thus:
\begin{equation}
\label{staticAR}
S(p) = \frac{1}{i\gamma\cdot p +M }
\to
\frac{g_{N,\Delta}^2}{i\gamma\cdot p +M } \,.
\end{equation}
Using this expedient, one obtains the results in the last column of Table~\ref{TabFEMasses}, where the parameters $g_{N,\Delta}$ were chosen in order to obtain values for the baryon masses which are consistent with estimates of the respective quark-core masses.

\begin{figure}[t]



\includegraphics[width=0.42\textwidth]{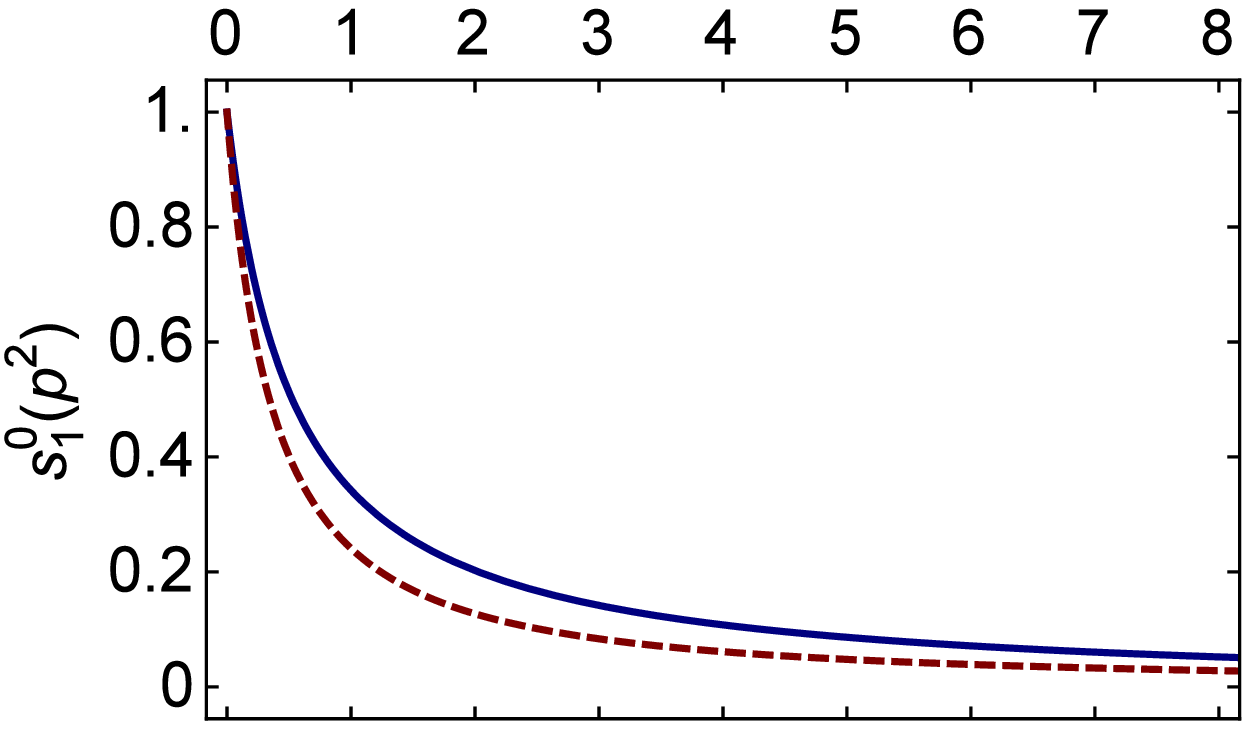}
\vspace*{-10ex}

\hspace*{-2.6ex}\includegraphics[width=0.435\textwidth]{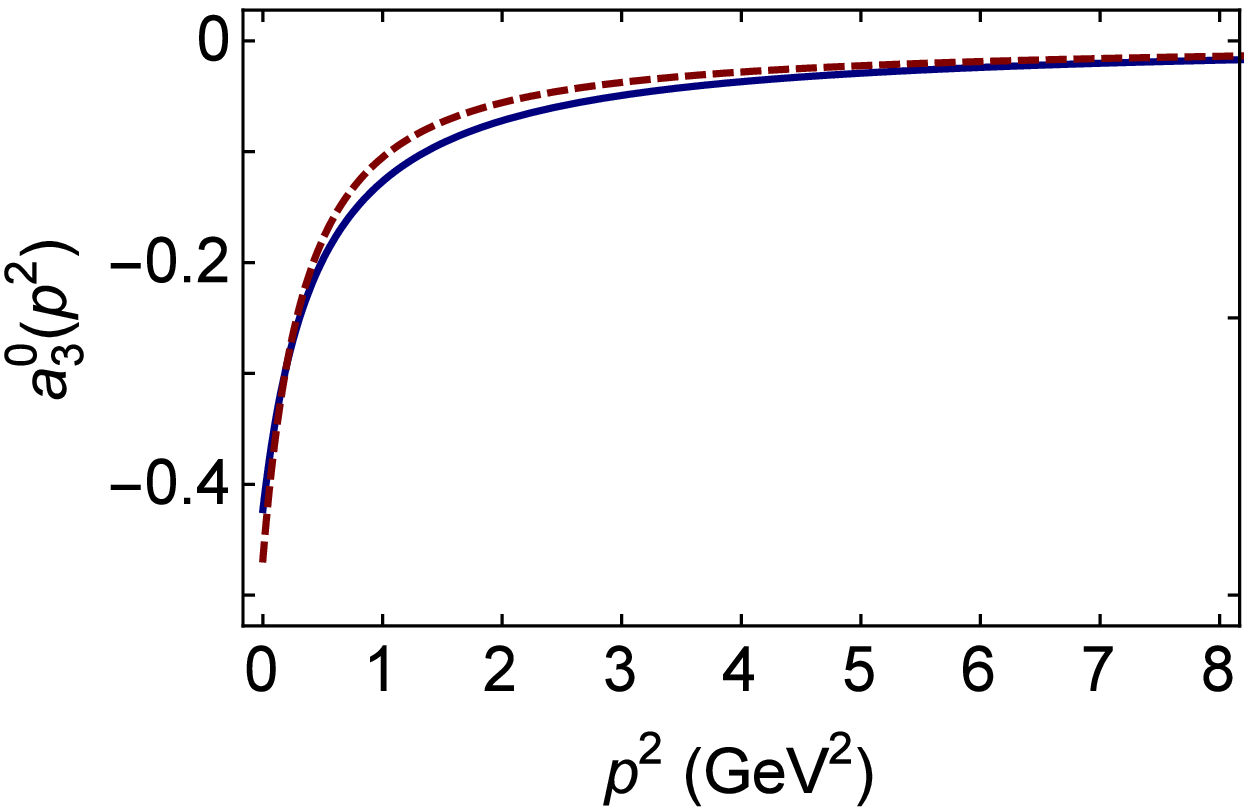}

\caption{\label{F2} \emph{Upper panel} -- Zeroth Chebyshev moment of the dominant amplitude in the scalar-diquark component of the nucleon's Faddeev amplitude, $s_1(p;P)$ in the notation of Eq.\,\eqref{SAStructure}, plotted for the two values of $g_N$ in Table~\ref{TabFEMasses}: solid (blue) $g_N=1.28$; red (dashed) $g_N=1.0$.  \emph{Lower panel} -- similar image for the dominant amplitude in the pseudovector-diquark component of the nucleon's Faddeev amplitude, $a_3(p;P)$.}
\end{figure}

It is important to determine whether the modification induced by Eq.\,\eqref{staticAR} has any effect on baryon internal structure.  That question can be answered by peering into the Faddeev amplitudes.  In the static approximation, the effect is modest, \emph{e.g}.\ including $g_N>1$ via the value in Table~\ref{TabFEMasses} increases the scalar-diquark component of the nucleon by 3\%.  The outcome is similar when the quark exchange is dynamical, as evident in Fig.\,\ref{F2}: with the normalisation fixed such that the zeroth Chebyshev moment of $s_1(p;P)$ is unity at $p^2=0$, the $p^2=0$ strength of the leading pseudovector component drops by $\lesssim 10$\% when $g_N=1\to1.28$ and the $p^2$-dependence of both moments is only affected modestly.

\begin{figure}[t]
\vspace*{-1ex}

\includegraphics[width=0.38\textwidth]{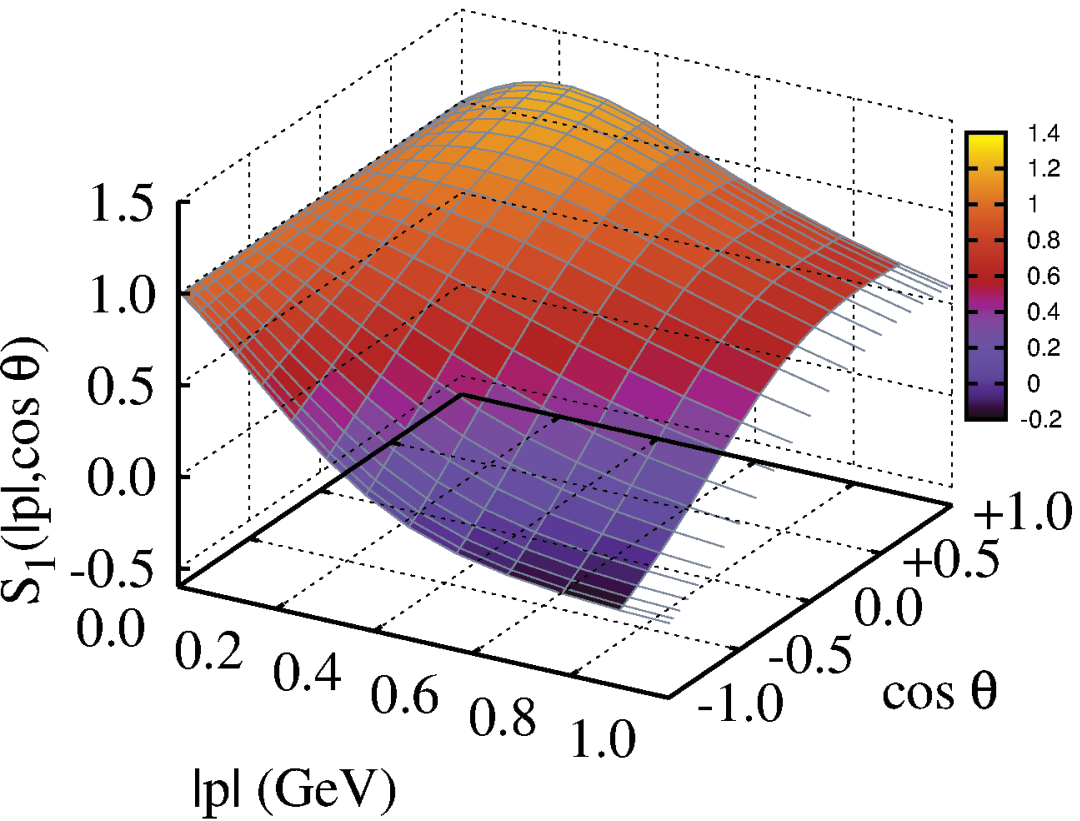}
\includegraphics[width=0.38\textwidth]{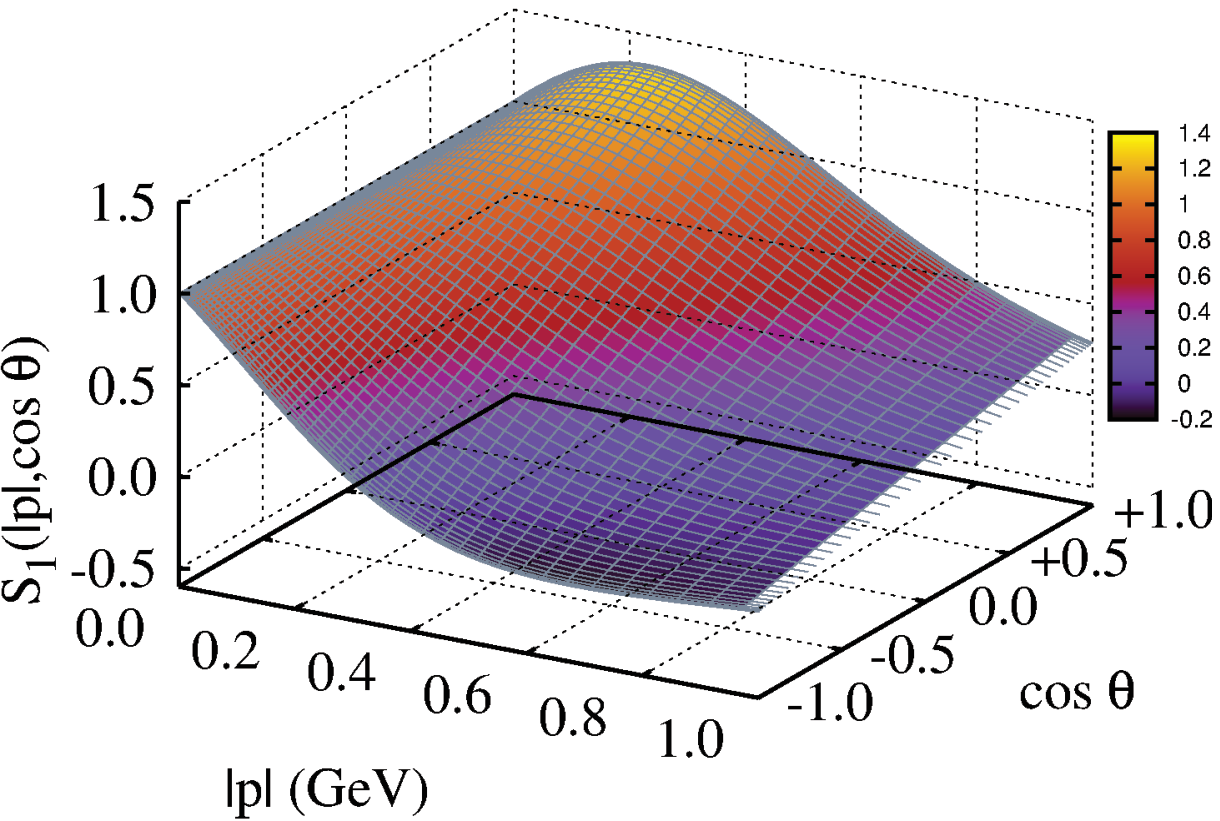}
\includegraphics[width=0.38\textwidth]{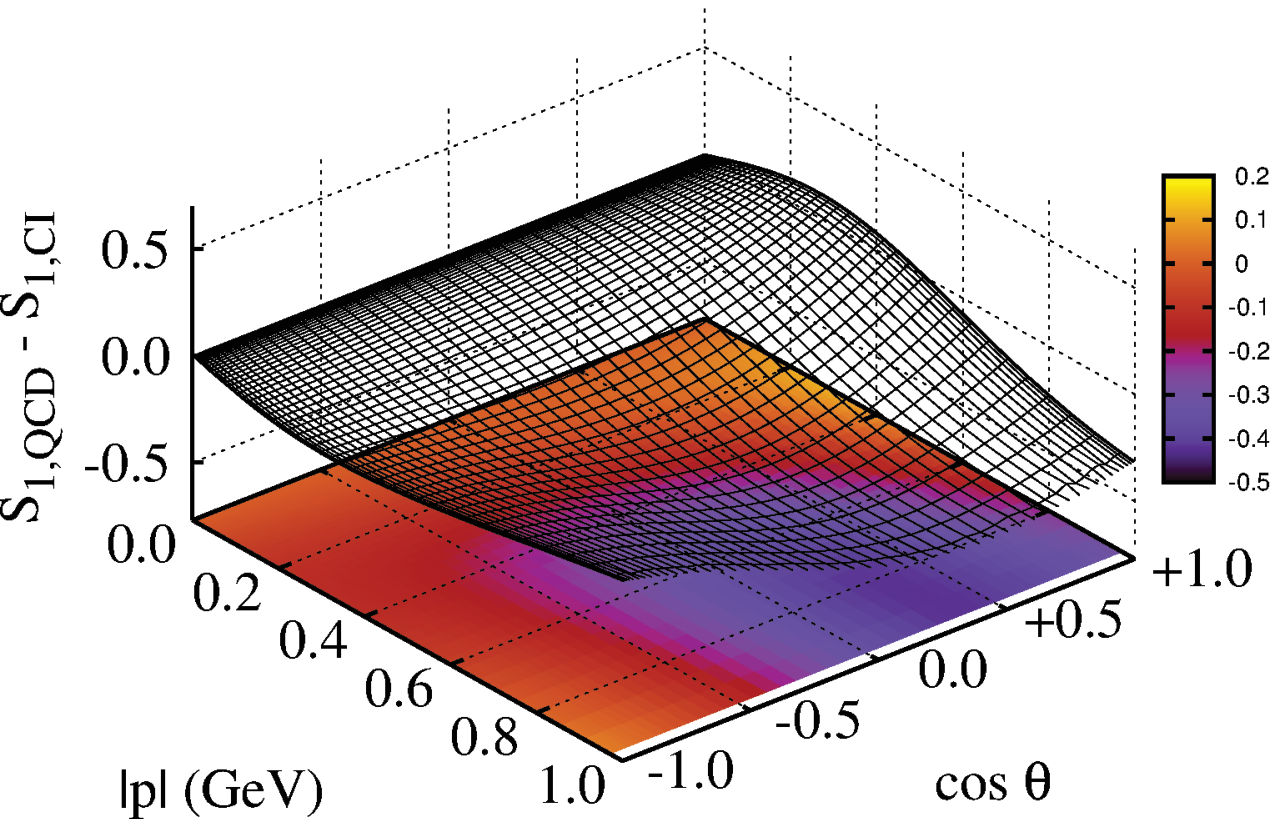}

\caption{\label{F3}
\emph{Top panel} -- Depiction of the complete spacetime dependence of the dominant piece in the nucleon's eight-component Poincar\'e-covariant Faddeev amplitude, $s_1(|p|,\cos\theta)$, computed herein using the contact interaction with dynamical quark exchange.
\emph{Middle panel} -- Same function computed using the QCD-kindred Faddeev equation kernel described in Refs.\,\cite{Segovia:2014aza, Segovia:2015ufa}, which differs from that used herein by using momentum-dependent dressed-quark masses and diquark Bethe-Salpeter correlation amplitudes.
\emph{Bottom panel} -- Difference between these two functions. }
\end{figure}

Figure~\ref{F2} also highlights a salient difference between our results and those obtained using the static approximation, Eq.\,\eqref{staticA}; namely, whereas the static approximation produces a momentum-independent result, our amplitude, obtained with dynamical quark exchange, exhibits strong momentum dependence: for $p^2\gtrsim 3 m_N^2$ the leading Chebyshev moments of the dominant scalar and pseudovector amplitudes depicted in Fig.\,\ref{F2} fall as $1/p^2$, up to $\ln p^2$-corrections.  Such behaviour is typical of two-body systems in quantum field theory; and here we have a quark-diquark system, with the diquark described by a momentum-independent Bethe-Salpeter correlation amplitude.

This outcome motivates a comparison, depicted in Fig.\,\ref{F3}, between the amplitude computed herein and that obtained using the  QCD-kindred Faddeev equation kernel described in Refs.\,\cite{Segovia:2014aza, Segovia:2015ufa}, which successfully unifies the description of nucleon, $\Delta$-baryon and Roper resonance properties and differs from ours by using momentum-dependent dressed-quark masses and diquark Bethe-Salpeter correlation amplitudes.  Whilst the spacetime dependence of the dominant piece in the nucleon's eight-component Poincar\'e-covariant Faddeev amplitude, $s_1(|p|,\cos\theta)$, computed using the contact interaction with dynamical quark exchange is quantitatively different from that obtained with the more realistic kernel, the functions nevertheless show marked qualitative similarities.  For example, the support for both is concentrated in the forward direction, $\cos\theta >0$, so that alignment of $p$ and $P$ is favoured; and the amplitude peaks at $(|p|\simeq M_N/6,\cos\theta=1)$, whereat $p_q \approx P/2 \approx p_d$ and hence the \emph{natural} relative momentum is zero.  In the antiparallel direction, $\cos\theta<0$, the support for both these functions is concentrated at $|p|=0$, \emph{i.e}.\ $p_q \approx P/3$, $p_d \approx 2P/3$.  Figure~\ref{F3} shows that using the static approximation in the contact-interaction Faddeev equation, in which case the analogous plot would simply depict the curve $s_1(|p|,\cos\theta)\equiv 1$, provides a description of nucleon structure that is badly flawed in connection with any probe sensitive to the nucleon interior; whereas implementation of dynamical quark exchange in the contact-interaction Faddeev equation yields a significant improvement in the description of a baryon's internal structure. %
Notwithstanding this, it should be borne in mind that when computed using a QCD-kindred kernel, the functions in Fig.\,\ref{F2} fall as $1/p^4$, up to $\ln p^2$-corrections, as one would expect of a three valence-body system.

\section{Probing the Nucleon}
\label{SecCurrents}
\subsection{Sigma Term}
Following Refs.\,\cite{Flambaum:2005kc, Holl:2005st}, it is straightforward to obtain the nucleon's scalar charge by using the Feynman-Hellmann theorem.  This method skirts the need to compute the canonical normalisation constant for the nucleon's Faddeev amplitude because one need only compute the variation of the nucleon's mass produced by a change in current-quark mass.  With the Faddeev equation kernel constructed as described in connection with Eqs.\,\eqref{njlgluon}, \eqref{RLvertex}, \eqref{staticAR}, one obtains
\begin{equation}
m_N(m) \stackrel{m\simeq 7\,{\rm MeV}}{=}
1.12 + 2.85 \, m + 8.54 \, m^2
\end{equation}
and hence
\begin{equation}
\left. \frac{\partial m_N}{\partial m}\right|^{\zeta_H}_{m=7\,{\rm MeV}} = 2.73 \,.
\end{equation}
This result is quoted at an hadronic scale appropriate to our formulation, \emph{viz}.\ $\zeta_H= 0.39 \pm 0.02\,$GeV (see Ref.\,\cite{Pitschmann:2014jxa}, Appendix E).  On the other hand, the product
\begin{equation}
\label{eqNsigmaF}
\sigma_N = m\, 2.73 = 19\,{\rm MeV}
\end{equation}
is independent of scale.   (Inclusion of meson-baryon loop effects is likely to increase the result in Eq.\,\eqref{eqNsigmaF} by roughly 7\,MeV \cite{Flambaum:2005kc}.)

The result in Eq.\,\eqref{eqNsigmaF} is approximately one-half that reported in computations using a QCD-derived interaction \cite{Flambaum:2005kc, Holl:2005st} and contemporary simulations of lattice-regularised QCD \cite{Shanahan:2012wh, Ren:2014vea}.  As explained elsewhere \cite{Pitschmann:2014jxa}, this mismatch exposes a defect of the contact interaction, \emph{viz}.\ it produces rigid, momentum-independent diquark Bethe-Salpeter amplitudes, an artefact which leads to a weaker $m$-dependence of the diquark (and hence nucleon) masses than is obtained with more realistic kernels.  Consequently, Eq.\,\eqref{eqNsigmaF} is an underestimate of $\sigma_N$.

\subsection{Canonical Normalisation}
\label{SecCN}
The $\sigma$-term is an exception.  Typically, one must calculate the normalisation before reporting a physical value for any observable.  The canonical normalisation constant for the Faddeev amplitude associated with an isospin multiplet is fixed by ensuring that the zero momentum-transfer $(Q^2=0)$ value of the electric form factor connected with a charged member of the multiplet is equal to the electric charge of that state \cite{Nakanishi:1969ph}.  In connection with the nucleon, whose Poincar\'e-covariant electromagnetic current may be written $(Q=P_f-P_i)$
\begin{equation}
\bar u(P_f)\big[ \gamma_\mu F_{1}(Q^2)+\tfrac{1}{2 m_N} \sigma_{\mu\nu} Q_\nu F_{2}(Q^2)\big] u(P_i)\,,
\label{NRcurrents}
\end{equation}
this means imposing \mbox{$F_1(Q^2=0)=1$} for the proton.

\begin{figure}[t]
\begin{minipage}[t]{0.5\textwidth}
\begin{minipage}[t]{0.45\textwidth}
\centerline{\includegraphics[width=0.90\textwidth]{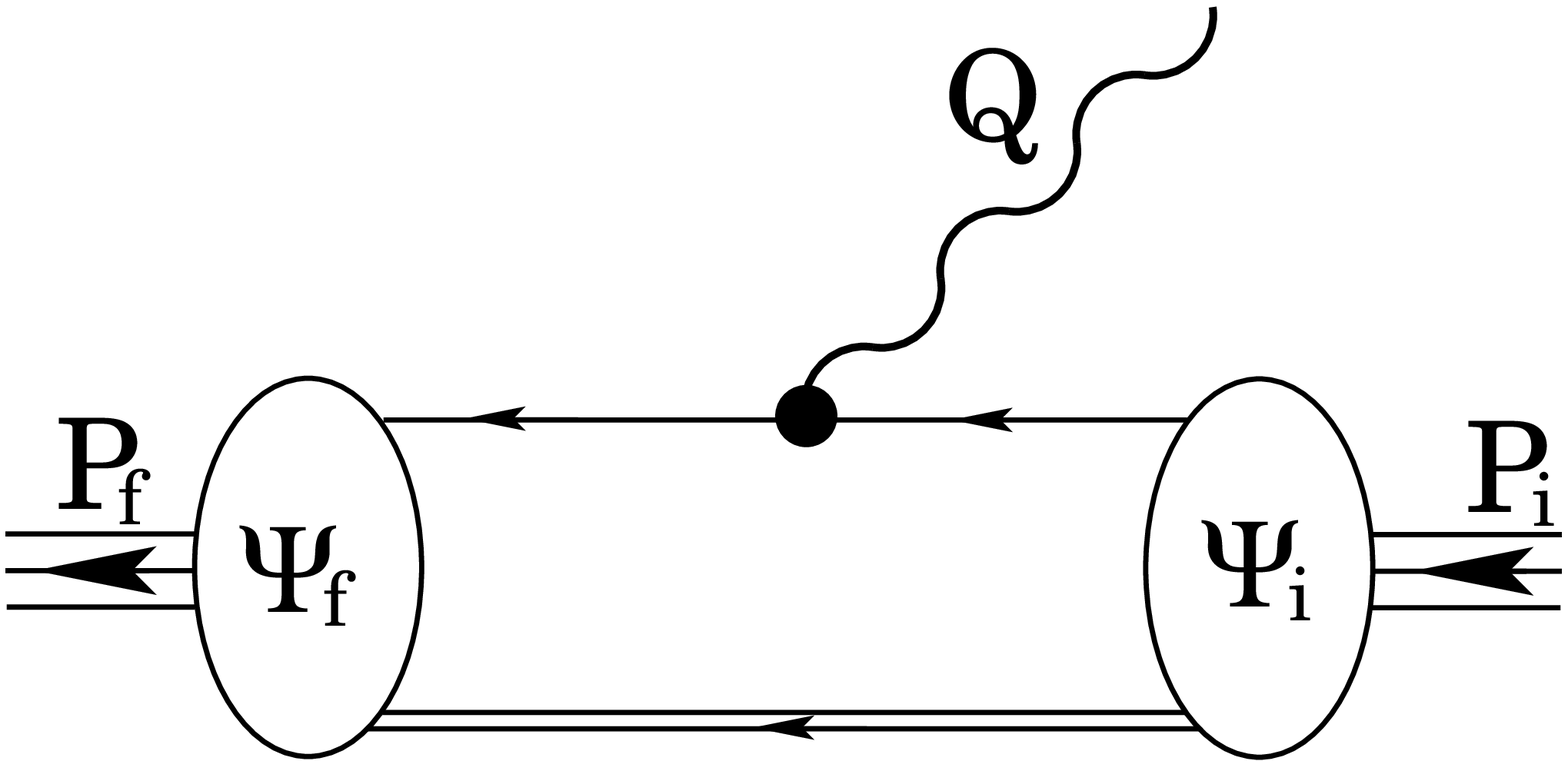}}
\end{minipage}
\begin{minipage}[t]{0.45\textwidth}
\centerline{\includegraphics[width=0.90\textwidth]{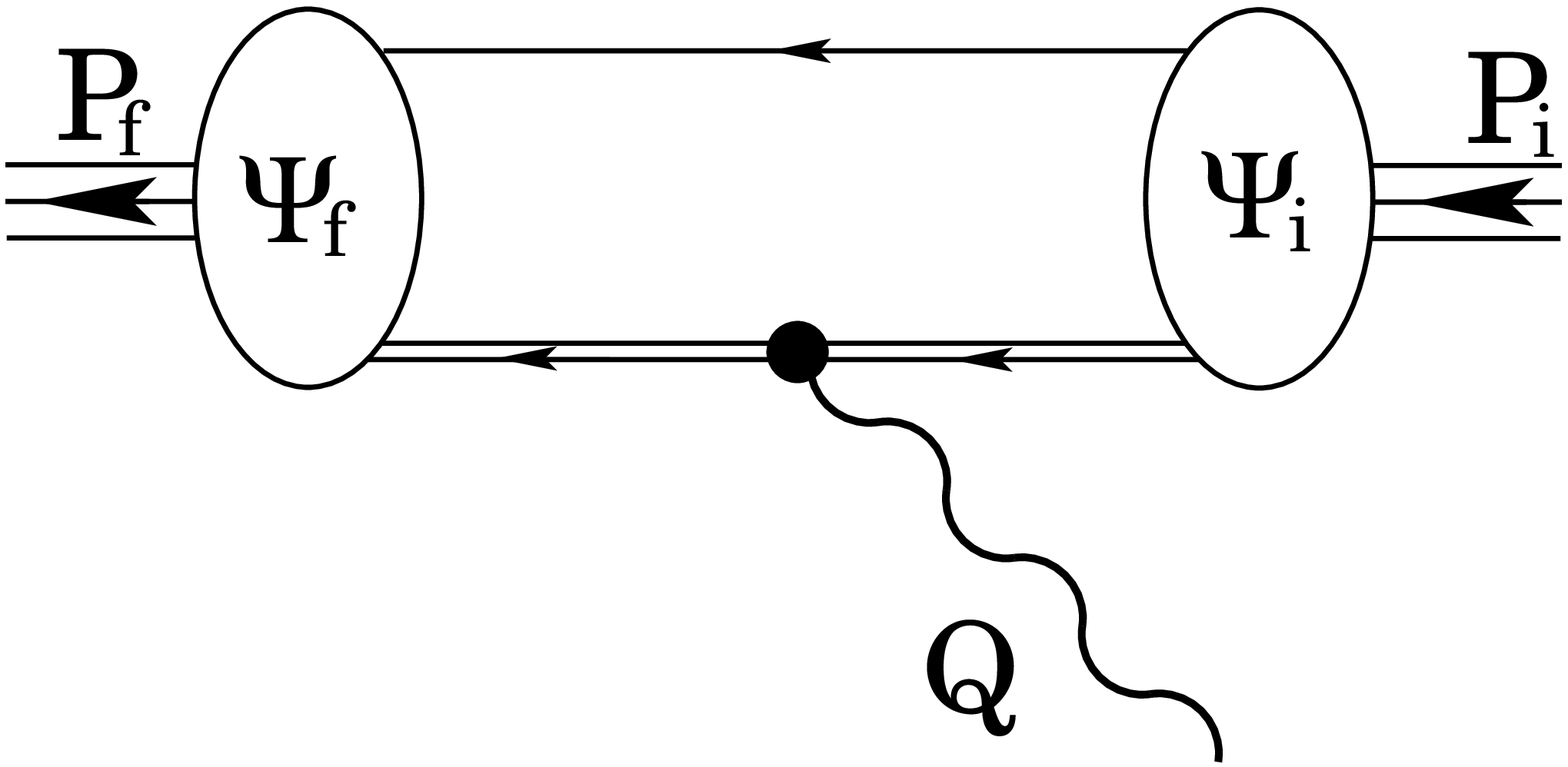}}
\end{minipage}\vspace*{3ex}

\begin{minipage}[t]{0.45\textwidth}
\centerline{\includegraphics[width=0.90\textwidth]{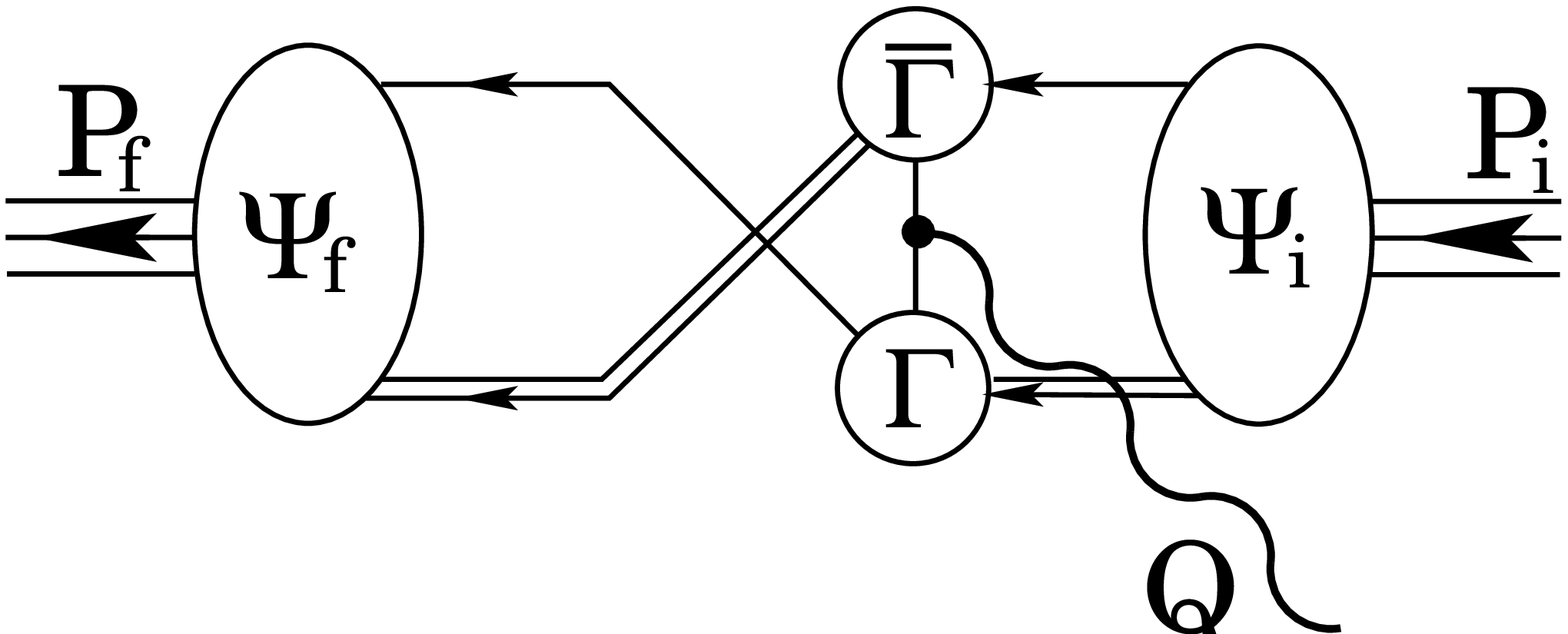}}
\end{minipage}
\begin{minipage}[t]{0.45\textwidth}
\centerline{\includegraphics[width=0.90\textwidth]{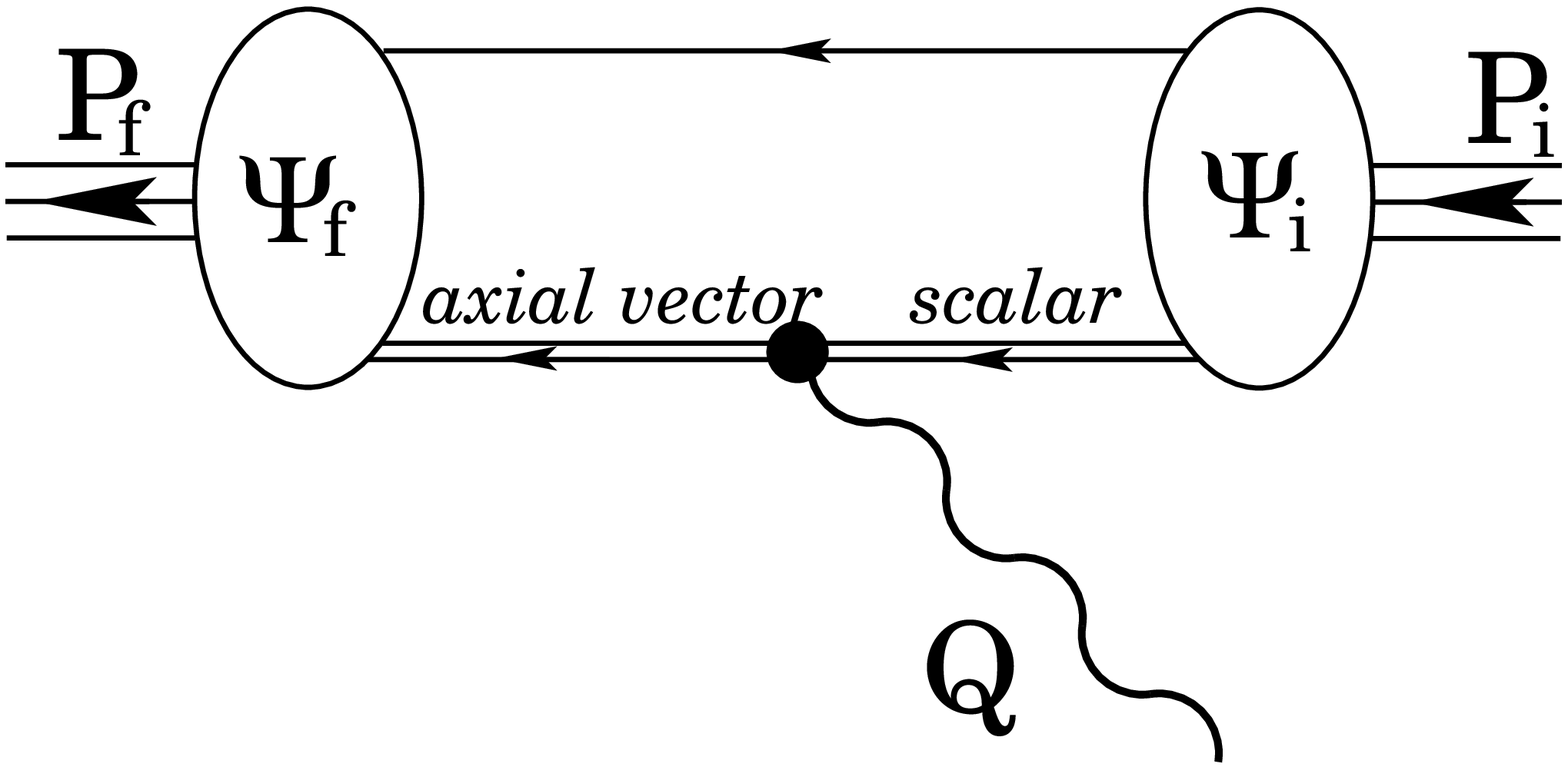}}
\end{minipage}
\end{minipage}
\caption{\label{CIcurrent} Vertex which ensures a conserved electromagnetic current for on-shell baryons described by the Faddeev amplitudes, $\Psi_{i,f}$, described in Sec.\,\protect\ref{SecFaddeev}.  As in Fig.\,\ref{figFaddeev}, the single line represents $S(p)$, the dressed-quark propagator, Sec.\,\protect\ref{SecGap}, and the double line, the diquark propagator, Sec.\,\protect\ref{SecDiquarks}; $\Gamma$ is the diquark Bethe-Salpeter amplitude, Sec.\,\protect\ref{SecDiquarks}; and the remaining vertices are described in Appendix\,\ref{NPVertex} -- the top-left image is Diagram~1; the top-right, Diagram~2; and so on, with the bottom-right image, Diagram~4.}
\end{figure}

The nucleon current is detailed in Ref.\,\cite{Oettel:1999gc}; and using a vector$\,\otimes\,$vector contact-interaction, that current may be described in terms of the four diagrams depicted in Fig.\,\ref{CIcurrent}.  (The so-called seagull terms vanish in our case because the diquark Bethe-Salpeter amplitudes are momentum-independent.)  Diagram~4 vanishes at $Q^2=0$ and hence does not contribute to the normalisation.  Its analogue does contribute to the tensor charge, however.  If one chooses to distinguish between quark flavours, the remaining diagrams produce fourteen distinct contributions: Diagrams~1 and 2, three each; and Diagram~3, eight terms.

Using the symbols $(0,\rightarrow,\uparrow)$ to denote diquark isospin labels -- $0$, scalar $[ud]$ diquark; $\rightarrow$, pseudovector $\{ud\}$ diquark; and $\uparrow$, pseudovector $\{uu\}$ diquark, then the distinct proton diagrams can be expressed as follows:
\begin{align}
\nonumber
& {\mathpzc C}^{1,2}_{00}\,, \; {\mathpzc C}^{1,2}_{\rightarrow \rightarrow}\,,\;
{\mathpzc C}^{1,2}_{\uparrow \uparrow}\,,\; \\
&
{\mathpzc C}^{3}_{00}\,,\; 
{\mathpzc C}^{3}_{0\rightarrow}\,,\;
{\mathpzc C}^{3}_{0\uparrow}\,,\;
{\mathpzc C}^{3}_{\rightarrow 0}\,,\; \\
\nonumber
&
{\mathpzc C}^{3}_{\rightarrow\rightarrow}\,,\;
{\mathpzc C}^{3}_{\rightarrow \uparrow}\,,\;
{\mathpzc C}^{3}_{\uparrow 0}\,, \;
{\mathpzc C}^{3}_{\uparrow \rightarrow}\,.
\end{align}
Here, \emph{e.g}.\ ${\mathpzc C}^{1}_{00}$ represents a $[ud]$ diquark in both the initial and final-state proton, with the probe striking a $u$-quark; and ${\mathpzc C}^{3}_{\uparrow \rightarrow}$ describes a $\{ud\}$ diquark in the initial-state, $\{uu\}$ in the final state, and the probe striking a $u$-quark exchanged between the two.  (The neutron diagrams are obtained by exchanging $u\leftrightarrow d$ so that ``$\uparrow$'' becomes ``$\downarrow$'' and the $\{uu\}$ diquark is replaced with the $\{dd\}$ correlation.)
Naturally, isospin symmetry reduces the number of truly independent computations to just seven:
\begin{align}
\nonumber
&
{\mathpzc C}^{1,2}_{00}\,,\;
{\mathpzc C}^{1,2}_{\uparrow \uparrow} \propto {\mathpzc C}^{1,2}_{\rightarrow \rightarrow}\,,\;
{\mathpzc C}^{3}_{00}\,,\; \\
&
{\mathpzc C}^{3}_{\uparrow 0} \propto {\mathpzc C}^{3}_{\rightarrow 0} \propto
{\mathpzc C}^{3}_{0\uparrow} \propto {\mathpzc C}^{3}_{0\rightarrow}\,,
\label{CNisospin}\\
\nonumber &
{\mathpzc C}^{3}_{\uparrow \rightarrow} \propto
{\mathpzc C}^{3}_{\rightarrow \uparrow} \propto
{\mathpzc C}^{3}_{\rightarrow\rightarrow}\,;
\end{align}
and current-conservation makes one of these redundant.  Our calculation of the normalisation is sketched in Appendix\,\ref{NPVertex}.

\subsection{Valence-quark Distributions}
It is now possible to exploit a connection between the $Q^2=0$ values of elastic form factors, \emph{i.e}.\ the Faddeev amplitude's normalisation, and the dimensionless structure functions of deep inelastic scattering at Bjorken-$x=:x_B=1$.  As remarked elsewhere \cite{Holt:2010vj, Wilson:2011aa, Roberts:2013mja, Segovia:2014aza}, whilst all familiar parton distribution functions (PDFs) vanish at $x_B=1$, ratios of any two need not; and, under DGLAP evolution \cite{Dokshitzer:1977, Gribov:1972, Lipatov:1974qm, Altarelli:1977}, the value of such a ratio is invariant.  Thus, \emph{e.g}.\ with $d_v(x_B)$, $u_v(x_B)$ the proton's $d$, $u$ valence-quark PDFs, the value of $\lim_{x_B\to 1} d_v(x_B)/u_v(x_B)$ is an unambiguous, scale invariant, nonperturbative feature of QCD.  It is therefore a keen discriminator between frameworks that claim to explain nucleon structure.  Furthermore, $x_B=1$ corresponds strictly to the situation in which the invariant mass of the hadronic final state is precisely that of the target, \emph{viz}.\ elastic scattering.  The structure functions inferred experimentally on the neighborhood $x_B\simeq 1$ are therefore determined theoretically by the target's elastic form factors.

In this connection, consider the current depicted in Fig.\,\ref{CIcurrent}.  Since diquarks are soft, the only contributions which survive at $x_B=1$ are those from Diagrams~1 and 3, in which the probe interacts with an isolated quark.  Each piece from these diagrams appears with a strength determined by the proton's Faddeev amplitude, which expresses the effect of weightings derived from both the amplitude's spacetime- and isospin-dependence.  These properties are expressed in Eq.\,\eqref{protoncharge}, which yields the following probabilities for finding a quark of flavour $f=u,d$ at $x_B=1$:
\begin{equation}
A_f = \frac{\partial e_p}{\partial e_f} \,,\quad P_f = A_f/(A_u+A_d)\,.
\end{equation}

Inserting computed values for each of the elements in Eq.\,\eqref{protoncharge}, one finds
\begin{equation}
\label{largexB}
P_u = 0.88\,,\; P_d=0.12\,, \quad \lim_{x_B\to 1} \frac{d_v(x_B)}{u_v(x_B)} = 0.14\,,
\end{equation}
which corresponds to $F_2^n/F_2^p = 0.38$ at $x_B=1$.  These results are collected in Table~\ref{tab:a}.  The value of $\lim_{x_B\to 1} d_v(x_B)/u_v(x_B)$ in Eq.\,\eqref{largexB} is 20\% smaller than that computed using the contact interaction in tandem with the static approximation \cite{Wilson:2011aa}.  In part this is because the static approximation kills Diagram~3 and that diagram favours hard $u$-quarks; but the reduction occurs mostly because removing the static approximation strengthens the Diagram~1 scalar-diquark contribution (hard $u$-quark only) relative to that from the heavier pseudovector-diquark.  (This can be seen via the discussion of Fig.\,\ref{F2}.)

\begin{table}[t]
\begin{center}
\caption{\label{tab:a}
Selected predictions for the $x_B=1$ value of the indicated quantities.
The DSE results are computed using the formulae in Eqs.\,\eqref{PDFratios}\,--\,\eqref{Pscalar}:
``realistic'' denotes results obtained with a sophisticated QCD-kindred Faddeev equation kernel \cite{Segovia:2014aza};
``contact-S'' are contact-interaction results obtained using a static approximation, described in connection with Eq.\,\eqref{staticA}; and
``contact-D'' are the results obtained herein, deriving from a Faddeev equation kernel with dynamical dressed-quark exchange.
The next four rows are, respectively, results drawn from Refs.\,\protect\cite{Close:1988br, Cloet:2005pp, Hughes:1999wr, Isgur:1998yb}: the row labelled $0_{[ud]}^+$-frozen reproduces results from a model in which a non-dynamical scalar-diquark is used to describe nucleon structure, \emph{i.e}. it is not based on a Faddeev equation and hence Diagrams~3 and 4 in Fig.\,\ref{CIcurrent} are absent.
The last row, labeled ``pQCD,'' expresses predictions made in Refs.\,\protect\cite{Farrar:1975yb,Brodsky:1994kg}, which are based on an SU$(6)$ spin-flavour wave function for the proton's valence-quarks and assume helicity conservation in their interaction with hard-photons (for reference, $3/7\approx 0.43$).
}
\begin{tabular*}
{\hsize}
{
l|@{\extracolsep{0ptplus1fil}}
l@{\extracolsep{0ptplus1fil}}
l@{\extracolsep{0ptplus1fil}}
l@{\extracolsep{0ptplus1fil}}
l@{\extracolsep{0ptplus1fil}}
l@{\extracolsep{0ptplus1fil}}
l@{\extracolsep{0ptplus1fil}}
l@{\extracolsep{0ptplus1fil}}}\hline
    & $\frac{F_2^n}{F_2^p}$ & $\frac{d}{u}$ & $\frac{\Delta d}{\Delta u}$
    & $\frac{\Delta u}{u}$ & $\frac{\Delta d}{d}$ & $A_1^n$ & $A_1^p$\\\hline
%
DSE-realistic \cite{Segovia:2014aza}
& $0.50$ & $0.29$ & $-0.12$ & 0.67 & $-0.29$ & 0.16 & 0.61 \\[0.5ex]
%
DSE-contact-S \cite{Roberts:2011wy}
& $0.41$ & $0.18$ & $-0.07$ & 0.88 & $-0.33$ & 0.34 & 0.88\\
DSE-contact-D
& $0.38$ & $0.14$ & $-0.05$ & 0.83 & $-0.33$ & 0.43 & 0.79\\\hline
$0_{[ud]}^+$-frozen & $\frac{1}{4}$ & 0 & $\phantom{-}$0 & 1 & $\phantom{-}$0 & 1 & 1 \\[0.5ex]
NJL & $0.43$ & $0.20$ & $-0.06$ & 0.80 & $-0.25$ & 0.35& 0.77 \\[0.5ex]
SU$(6)$ & $\frac{2}{3}$ & $\frac{1}{2}$ & $-\frac{1}{4}$ & $\frac{2}{3}$ & $-\frac{1}{3}$ & 0 & $\frac{5}{9}$ \\[0.5ex]
CQM & $\frac{1}{4}$ & 0 &$\phantom{-}$0 & 1 & $-\frac{1}{3}$ & 1 & 1 \\[0.5ex]
pQCD & $\frac{3}{7}$ & $\frac{1}{5}$ & $\phantom{-}\frac{1}{5}$ & 1 & $\phantom{-}$1 & 1 & 1 \\\hline
\end{tabular*}
\end{center}
\end{table}

The observations that open this subsection have also been exploited \cite{Roberts:2013mja} in order to deduce a collection of simple formulae, expressed in terms of the diquark appearance and mixing probabilities, from which one may compute ratios of longitudinal-spin-dependent $u$- and $d$-quark parton distribution functions on the domain $x_B\simeq 1$:
\begin{subequations}
\label{PDFratios}
\begin{align}
\label{dvuvF1result}
%
\quad A_1^n & = \frac{
4 (P_{d_\uparrow}-P_{d_\downarrow}) + (P_{u_\uparrow}-P_{u_\downarrow})}
{4(P_{d_\uparrow}+P_{d_\downarrow}) + (P_{u_\uparrow}+P_{u_\downarrow})}\,,\\
%
%
A_1^p &= \frac{
4 (P_{u_\uparrow}-P_{u_\downarrow}) + (P_{d_\uparrow}-P_{d_\downarrow})}
{4 (P_{u_\uparrow}+P_{u_\downarrow}) +(P_{d_\uparrow}+P_{d_\downarrow})}\,,
\end{align}
\end{subequations}
where the probabilities for the different quark flavours to have helicity aligned ($\uparrow$) or opposite ($\downarrow$) to that of the proton are:
{\allowdisplaybreaks
\begin{subequations}
\label{proball}
\begin{align}
P_{u_\uparrow} & = P^{p,s}_{u_\uparrow} + \frac{1}{9} P^{p,a} + \frac{1}{3} P^{p,m} \\
%
%
P_{u_\downarrow} & = P^{p,s}_{u_\downarrow} + \frac{2}{9} P^{p,a} + \frac{1}{3} P^{p,m},\\
%
%
P_{d_\uparrow} & = \frac{2}{9} P^{p,a} + \frac{1}{6} P^{p,m}, \\
P_{d_\downarrow} & = \frac{4}{9} P^{p,a} + \frac{1}{6} P^{p,m}\,.
\end{align}
\end{subequations}}
\hspace*{-0.5\parindent}The first line of Eq.\,\eqref{proball} can be understood once one understands that $P^{p,s}$ is the probability for finding a $u$-quark bystander in association with a scalar $[ud]$-diquark correlation in the proton.  Owing to Poincar\'e covariance, this term expresses a sum of quark-diquark angular momentum $L^{u[ud]}=0$ and $L^{u[ud]}=1$ correlations within the nucleon.  With $L^{u[ud]}=0$, the bystander quark carries all the nucleon's spin.  On the other hand, the $L^{u[ud]}=1$ correlation contributes to both the parallel and antiparallel alignment probabilities of the bystander quark: $2 [ud]_{L_z^{u[ud]}=1} u_{\downarrow} \oplus [ud]_{L_z^{u[ud]}=0} u_{\uparrow}$.  The relative strength of these terms is fixed by solving the Faddeev equation and expressed thereafter in the Faddeev amplitude: $\Psi_{0^+} \sim \psi_{L=0} + \psi_{L=1}$, so that, converting the amplitude to probabilities,
\begin{equation}
\label{Pscalar}
\begin{array}{ll}
P^{p,s} & = P^{p,s}_{u_\uparrow} + P^{p,s}_{u_\downarrow},  \\
P^{p,s}_{u_\uparrow} & = \psi_{L=0}^2 + 2 \psi_{L=0} \psi_{L=1}+ \frac{1}{3} \psi_{L=1}^2, \\
P^{p,s}_{u_\downarrow} & =  \frac{2}{3} \psi_{L=1}^2.\;
\end{array}
\end{equation}
With the Faddeev equation used herein, $P^{p,s}=0.82$, $\psi_{L=0}=0.67$, $\psi_{L=1}=0.24$ \emph{cf}.\ $\psi_{L=0}=0.55$, $\psi_{L=1}=0.22$ in Ref.\,\cite{Segovia:2014aza} and $\psi_{L=0}=0.88$, $\psi_{L=1}=0$ in Ref.\,\cite{Wilson:2011aa}.

The other two quantities in Eqs.\,\eqref{proball} are $P^{p,a}$, $P^{p,m}$, which, respectively, gauge the probability that the photon interacts with an axial-vector diquark component of the nucleon or induces a transition between diquark components of the incoming and outgoing nucleon.  Our dynamical dressed-quark exchange Faddeev kernel generates $P^{p,a}=0.18$ and $P^{p,m}\approx 0$; and we list results for numerous ratios in Table~\ref{tab:a}.  (As remarked above, the result $P^{p,m}\approx 0$ owes to the generally small magnitude of each Diagram~3 term and interference between their contributions, see Table~\ref{NormalisationContributions}.)

It is worth reiterating that the results in Table~\ref{tab:a} highlight that no single ratio is capable of completely distinguishing between distinct pictures of nucleon structure.  Conversely, they show that a comparison between experiment and different predictions for the combination of \emph{all} tabulated quantities provides a very effective means of discriminating between competing descriptions \cite{Roberts:2013mja}.

\subsection{Tensor Charges}
With the normalisation computed, one may also readily calculate the proton's tensor charges, which are defined via $(q=u,d)$:
\begin{equation}
\langle P(p,\sigma)|\bar{q}\sigma_{\mu\nu} q | P(p,\sigma)\rangle = \delta_T q\, \bar{u}(p,\sigma)\sigma_{\mu\nu} u(p,\sigma)\,,
\end{equation}
where $|P(p,\sigma)\rangle$ is a state vector describing a proton with momentum $p$ and spin $\sigma$.\footnote{In the isospin symmetric limit: $\delta_T^p u :=\delta_T u = \delta_T^n d$, $\delta_T^p d :=\delta_T d = \delta_T^n u$, where the superscripts denote the hadron in which the indicated valence-quark resides.}  The derived isoscalar and isovector tensor charges are:
\begin{equation}
\label{TensorIso}
  g_T^{(0)} = \delta_T u + \delta_T d\,, \;
  g_T^{(1)} = \delta_T u - \delta_T d\,.
\end{equation}
Importantly, the tensor charges are scale-dependent quantities, as explained, for instance, in Appendix\,F of Ref.\,\cite{Pitschmann:2014jxa}.  The values decrease uniformly as the resolving scale is increased.  We compute the results at $\zeta_H$ and use one-loop evolution equations in order to also report values at $\zeta_2:=2\,$GeV.


The nucleon's tensor interaction is qualitatively identical to the photon-nucleon interaction depicted in Fig.\,\ref{CIcurrent}.  If one distinguishes between quark flavours in this case, there are sixteen distinct contributions to each tensor charge because Diagram~4 is nonzero.  Using isospin symmetry, this tally is reduced to only eight.  The analysis is illustrated via consideration of Diagram~4 in Appendix\,\ref{PTensor}.  Notably,
\begin{equation}
{\mathpzc C}^{4 T}_{\rightarrow 0} = {\mathpzc C}^{4 T}_{0 \rightarrow}
\end{equation}
and ${\mathpzc C}^{2 T}_{00}\equiv 0$ because a scalar correlation cannot possess a tensor charge.

Our computed results for the proton's tensor charge at the hadronic scale $\zeta_H$ are summarised in the upper panel of Table~\ref{TensorContributions}.
It is natural to compare the listed values with those in Ref.\,\cite{Pitschmann:2014jxa}, obtained using the contact interaction and the static approximation, and listed in Table~C3 therein.  To facilitate that comparison, we note that Diagram~1 herein equates to the sum of Diagrams 1 and 2 therein, Diagram~2 herein is the sum of Diagrams 3 and 4 therein, Diagram 3 is absent in Ref.\,\cite{Pitschmann:2014jxa} owing to the static approximation, and our Diagram~4 equates to the sum of Diagrams 5 and 6 therein.
The net results of both calculations are semi-quantitatively similar.  However, there are some line-item discrepancies.   We have a nonzero result for Diagram~3, which adds to both $\delta u$,  $\delta d$, whereas this contribution is naturally absent when the static approximation is used.  There is also a difference between the Diagram~2 contributions.  This owes to a combination of the inadvertent omission in Ref.\,\cite{Pitschmann:2014jxa} of a factor of two in the probe-diquark vertices, which we have restored, with the corrected results listed in the lower panel of Table~\ref{TensorContributions}, combined with our use of Eqs.\,\eqref{onshellreplacements} in simplifying computation of the currents.

\begin{table}[t]
\caption{\label{TensorContributions}
\emph{Upper panel} --
Proton tensor charges evaluated at the model scale: $\zeta_H=0.39\pm0.02\,$GeV, partitioned according to contributions from the diagrams in Fig.\,\ref{CIcurrent} and summed to provide the complete result. 
\emph{Lower panel} --
Same quantities evaluated using the static approximation.  The listed values serve as an update of the results in Ref.\,\cite{Pitschmann:2014jxa}.
}
\begin{center}
\begin{tabular*}
{\hsize}
{
l@{\extracolsep{0ptplus1fil}}
c@{\extracolsep{0ptplus1fil}}
c@{\extracolsep{0ptplus1fil}}
c@{\extracolsep{0ptplus1fil}}
c@{\extracolsep{0ptplus1fil}}
c@{\extracolsep{0ptplus1fil}}}\hline
Dynamic          & $\delta u$ & $\delta d$ & $g_T^{(0)}$ & $g_T^{(1)}$ \\\hline
Diagram~1 & $\phantom{-}0.72$ & $-0.039$ & $\phantom{-}0.69\phantom{0}$ & $0.76$  \\
Diagram~2 & $\phantom{-}0.14$ & $\phantom{-}0.027$& $\phantom{-}0.16\phantom{0}$ & $0.11$ \\
Diagram~3 & $\phantom{-}0.12$ & $-0.050$ & $\phantom{-}0.075$ & $0.17$ \\
Diagram~4 & $-0.19$ & $-0.19\phantom{0}$ & $-0.38\phantom{0}$ & $0\phantom{.17}$ \\\hline
Total     & $\phantom{-}0.79$ & $-0.25\phantom{0}$ & $\phantom{-}0.54\phantom{0}$ & $1.05$ \\\hline
\end{tabular*}
\end{center}
\begin{center}
\begin{tabular*}
{\hsize}
{
l@{\extracolsep{0ptplus1fil}}
c@{\extracolsep{0ptplus1fil}}
c@{\extracolsep{0ptplus1fil}}
c@{\extracolsep{0ptplus1fil}}
c@{\extracolsep{0ptplus1fil}}
c@{\extracolsep{0ptplus1fil}}}\hline
Static          & $\delta u$ & $\delta d$ & $g_T^{(0)}$ & $g_T^{(1)}$ \\\hline
Diagram~1 & $\phantom{-}0.56$ & $-0.036$ & $\phantom{-}0.53\phantom{0}$ & $0.60$  \\
Diagram~2 & $\phantom{-}0.29$ & $\phantom{-}0.059$& $\phantom{-}0.35\phantom{0}$ & $0.23$ \\
Diagram~3 & $\phantom{-}0$ & $\phantom{-}0$ & $\phantom{-}0$ & $\phantom{-}0$ \\
Diagram~4 & $-0.25$ & $-0.25\phantom{0}$ & $-0.50\phantom{0}$ & $\phantom{-}0$ \\\hline
Total     & $\phantom{-}0.61$ & $-0.23\phantom{0}$ & $\phantom{-}0.38\phantom{0}$ & $0.83$ \\\hline
\end{tabular*}
\end{center}
\end{table}

It is worth analysing the microscopic origin of the proton's tensor charges in our calculation.  The dominant contribution to $\delta_T u$ arises from Diagram~1: tensor probe interacting with a dressed $u$-quark when a scalar diquark is the bystander.  The next largest piece is produced by Diagram~4, in which the tensor probe excites a transition between the scalar and pseudovector diquarks; but this is largely cancelled by the sum of Diagram~2 (tensor probe interacting with the pseudovector diquark with a dressed-quark spectator) and Diagram~3 (two-loop diagrams in which the tensor probe interacts with a dressed-quark ``in-flight'').  Qualitatively equivalent interference was seen in Ref.\,\cite{Pitschmann:2014jxa}, so that one may still conclude that $\delta_T u$ directly probes the strength of DCSB and hence the strong interaction at infrared momenta.  Owing to analogous interference between Diagrams~1-3, Diagram~4 in Fig.\,\ref{CIcurrent} is responsible for the bulk of $\delta_T d$.

Notably herein, in contrast to Ref.\,\cite{Pitschmann:2014jxa}, $\delta_T d\neq 0$ even in the absence of pseudovector diquark correlations: owing to Diagram~3, the tensor probe can interact with a dressed $d$-quark exchanged during the breakup and reformation of a scalar diquark.  However, as already noted, the contribution is small, \emph{viz}.\ this term produces just 20\% of the Diagram~3-$\delta d$ entry in the upper panel of Table~\ref{TensorContributions} and all other entries would be zero in the absence of pseudovector diquarks.  The magnitude of $\delta_T d$ may therefore still be interpreted as a measure of the strength of pseudovector diquark correlations in the proton.

In ascribing an error to our final result for the tensor charges, we follow the reasoning in Ref.\,\cite{Pitschmann:2014jxa}.  Namely, since one generally finds that systematic treatments of the contact interaction yield results for low-momentum-transfer observables which are practically indistinguishable from those produced by RL studies that employ more sophisticated interactions \cite{GutierrezGuerrero:2010md, Roberts:2010rn, Roberts:2011cf, Roberts:2011wy, Wilson:2011aa, Chen:2012qr, Chen:2012txa, Pitschmann:2012by, Wang:2013wk, Segovia:2013rca, Segovia:2013uga}; and analyses of hadron physics observables using the RL truncation and one-loop QCD renormalisation-group-improved (RGI) kernels for the gap and bound-state equations produce results that are normally within 15\% of the experimental value \cite{Roberts:2007jh}, we therefore attach a relative error of 15\% to the results in Table~\ref{TensorContributions}.  Hence, our predictions are:
\begin{equation}
\label{errorResults}
\begin{array}{l|cccc}
&\delta_T u & \delta_T d & g_T^{(0)} & g_T^{(1)}\\
\zeta_H \approx M & 0.79(12) & -0.25(4) & 0.54(8) & 1.05(16)
\end{array}\,.
\end{equation}

Given that our computed value of the proton's $\sigma$-term is too small by a factor of roughly two, one might be concerned by the size of our error assignment.  This concern can be allayed by first noting that the small size of $\sigma_p$ can be tracked directly to an underestimate of the diquarks' $\sigma$-terms \cite{Pitschmann:2014jxa}.  These quantities measure the rate-of-change of a ``charge'' (the diquark mass, in this case) associated with variations in an external source.  Although those rates-of-change are underestimated, the masses are not.  Thus, whilst our framework might produce rates-of-change for the tensor charges that are too small, the values of the tensor charges themselves should be accurate within the usual error associated with rainbow-ladder truncation.

The results in Eq.\,\eqref{errorResults} are quoted at the model scale.  In order to make a sensible comparison with estimates obtained in modern simulations of lattice-regularised QCD, those results must be evolved to $\zeta_2=2\,$GeV.  We therefore list here the results obtained under leading-order evolution to $\zeta_2=2\,$GeV, obtained via multiplication by the factor $0.794$, explained and computed in Appendix~F of Ref.\,\cite{Pitschmann:2014jxa}:
\begin{equation}
\label{tensorz2}
\begin{array}{l|cccc}
&\delta_T u & \delta_T d & g_T^{(0)} & g_T^{(1)}\\
\zeta_2 & 0.63(9) & -0.20(3) & 0.43(6) & 0.83(12)
\end{array}\,.
\end{equation}
We use the one-loop expression owing to the simplicity of our framework.  Employing next-to-leading-order evolution leads simply to a 25\% increase in $\zeta_{\rm H}$ with no material phenomenological differences.

\begin{figure}[t!]
\includegraphics[width=1.0\linewidth]{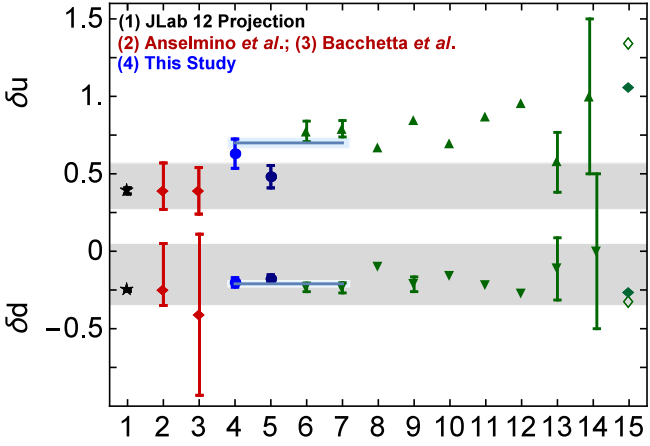}
\caption{Flavour separation of the proton's tensor charge:
``1'' -- illustration of anticipated accuracy in planned JLab experiment \cite{Gao:2010av}, with central values based on the analysis in Ref.\,\cite{Anselmino:2013vqa};
``2'' -- results drawn from Ref.\,\cite{Anselmino:2013vqa};
``3'' phenomenological estimate in Ref.\,\cite{Radici:2015mwa}
``4'' -- prediction herein, Eq.\,\eqref{tensorz2};
``5'' -- corrected results from Ref.\,\cite{Pitschmann:2014jxa}, drawn from the lower panel of Table~\ref{TensorContributions} and evolved to $\zeta_2$;
``6-14'' -- estimates from Refs.\,\cite{
Bhattacharya:2015wna, 
Abdel-Rehim:2015owa, 
Hecht:2001ry, 
Gockeler:2005cj, 
Cloet:2007em, 
Pasquini:2006iv, 
Wakamatsu:2007nc, 
Gamberg:2001qc, 
He:1994gz}, respectively.  %
The bands drawn from ``4''--``7'' are described in connection with Eq.\,\eqref{DSElQCD}.
By way of context, we note that were the proton a weakly-interacting collection of three massive valence-quarks, then \cite{He:1994gz} the quark axial and tensor charges are identical, so that $\delta_T u =4/3$ and $\delta_T d=-1/3$ at the model scale.  These values are located at ``15''.  If one assumes that $\zeta_H$ may reasonably be assigned as the scale of that model, then those values evolve to $\delta_T u =1.06$ and $\delta_T d=-0.26$, which we also plot at ``15''.
\label{TensorAll}}
\end{figure}

The predictions in Eq.\,\eqref{tensorz2} are compared in Fig.\,\ref{TensorAll} with phenomenological analyses \cite{Anselmino:2013vqa, Radici:2015mwa} that benchmark a proposed JLab experiment \cite{Gao:2010av} and results obtained using numerous other methods
\cite{
Bhattacharya:2015wna, 
Abdel-Rehim:2015owa, 
Hecht:2001ry, 
Gockeler:2005cj, 
Cloet:2007em, 
Pasquini:2006iv, 
Wakamatsu:2007nc, 
Gamberg:2001qc, 
He:1994gz}.
Evidently, our predictions match, within errors, the most recent results obtained using lattice-QCD \cite{Bhattacharya:2015wna, Abdel-Rehim:2015owa}, which appear at ``6'' and ``7'' in Fig.\,\ref{TensorAll}.  A weighted combination of the most recent DSE- and lattice-QCD results yields the following estimates, drawn in Fig.\,\ref{TensorAll}:
\begin{equation}
\label{DSElQCD}
\delta_T u =0.70 \pm 0.03\,, \quad \delta_T d = -0.21 \pm 0.01\,.
\end{equation}

Another interesting point is highlighted by a comparison between our predictions and the values obtained when the proton is considered to be a weakly-interacting collection of three massive valence-quarks described by an SU$(4)$-symmetric spin-flavour wave function \cite{He:1994gz}: $\delta_T^{{\rm SU}(4)} u =2 e_u$ and $\delta_T^{{\rm SU}(4)} d=e_d$ \emph{cf}.\ our results, Eq.\,\eqref{errorResults}, $\delta_T u = 0.59 (2 e_u)$, $\delta_T d = 0.75(e_d)$.  It is thus apparent that the presence of diquark correlations in the proton Faddeev amplitude materially suppresses the magnitude of the tensor charge associated with each valence quark whilst simultaneously increasing the ratio $\delta_T d/\delta_T u$ by approximately 30\%.

\subsection{Electric Dipole Moments}
\label{SecEDM}
In typical extensions of the Standard Model, quarks acquire an EDM \cite{Pospelov:2005pr, RamseyMusolf:2006vr}, \emph{i.e}.\ an interaction with the photon that proceeds via a current of the form:
\begin{equation}
\tilde d_q \, q \gamma_5 \sigma_{\mu\nu} q\,,
\end{equation}
where $\tilde d_q$ is the quark's EDM and here we consider $q=u,d$.  The EDM of a proton containing quarks which interact in this way is defined as follows:
\begin{equation}
\langle P(p,\sigma) | \mathcal J_{\mu\nu}^\text{EDM} | P(p,\sigma)\rangle
= \tilde d_p\, \bar {\mathpzc u}(p,\sigma)\gamma_5\sigma_{\mu\nu} {\mathpzc u}(p,\sigma)\,,
\end{equation}
where
\begin{equation}
 \mathcal J_{\mu\nu}^\text{EDM}(x) = \tilde d_u\,\bar u(x)\gamma_5\sigma_{\mu\nu}u(x) + \tilde d_d\,\bar d(x)\gamma_5\sigma_{\mu\nu}d(x)\,.
\end{equation}

At this point, using a simple Dirac-matrix identity:
\begin{equation}
\label{Epsilon}
\gamma_5\sigma_{\mu\nu} = \tfrac{1}{2} \varepsilon_{\mu\nu\alpha\beta} \sigma_{\alpha\beta}\,,
\end{equation}
one can write
\begin{equation}
\mathcal J_{\mu\nu}^\text{EDM}
= \tfrac{1}{2} \varepsilon_{\mu\nu\alpha\beta}
\left[ \tilde d_u\,\bar u \sigma_{\alpha\beta}u  + \tilde d_d\,\bar d\sigma_{\alpha\beta}d\right].
\end{equation}
It follows that
\begin{align}
&\langle P(p,\sigma) | \mathcal J_{\mu\nu}^\text{EDM} | P(p,\sigma)\rangle \nonumber\\
&= \left[
\tilde d_u \, \delta_T u\,
+ \tilde d_d \, \delta_T d\,
\right]\bar {\mathpzc u}(p,\sigma)\gamma_5\sigma_{\mu\nu}{\mathpzc u}(p,\sigma) \,;
\end{align}
namely \cite{Pitschmann:2014jxa}, the quark-EDM contribution to a proton's EDM is completely determined once the proton's tensor charges are known:
\begin{equation}
\tilde d_p = \tilde d_u \, \delta_T u\, + \tilde d_d \, \delta_T d\,.
\end{equation}
With emerging experimental techniques, it is possible to place competitive upper-limits on the proton's EDM using storage rings in which polarized particles are exposed to an electric field \cite{Pretz:2013us}.

An analogous result for the neutron is readily inferred.  In the limit of isospin symmetry,
\begin{align}
\langle N(p,\sigma)|\bar u\sigma_{\mu\nu}u|N(p,\sigma)\rangle &= \langle P(p,\sigma)|\bar d\sigma_{\mu\nu}d|P(p,\sigma)\rangle\,, \nonumber\\
\langle N(p,\sigma)|\bar d\sigma_{\mu\nu}d|N(p,\sigma)\rangle &= \langle P(p,\sigma)|\bar u\sigma_{\mu\nu}u|P(p,\sigma)\rangle\,;
\end{align}
and hence
$\tilde d_n = \tilde d_u \, \delta_T d\, + \tilde d_d \, \delta_T u$.
Using the results in Eq.\,\eqref{errorResults}, we therefore have
\begin{equation}
\label{dndp}
\tilde d_n = -0.25\,\tilde d_u + 0.79\,\tilde d_d\,,\;
\tilde d_p = 0.79\,\tilde d_u - 0.25\,\tilde d_d\,.
\end{equation}

It is worth contrasting Eqs.\,\eqref{dndp} with the results one would obtain by assuming that the nucleon is simply a collection of three massive valence-quarks described by an SU$(4)$-symmetric spin-flavour wave function.  Then, by analogy with magnetic moment computations, a procedure also made valid by Eq.\,\eqref{Epsilon}:
\begin{equation}
\tilde d_n = -\tfrac{1}{3} \,\tilde d_u + \tfrac{4}{3} \,\tilde d_d\,,\;
\tilde d_p = \tfrac{4}{3} \,\tilde d_u - \tfrac{1}{3} \,\tilde d_d\,,
\end{equation}
values which are roughly 50\% larger than ours.

The impact of our tensor-charge predictions on beyond-Standard-Model phenomenology may be elucidated, \emph{e.g}.\ by following the analysis in Refs.\,\cite{Bhattacharya:2011qm, Dekens:2014jka}.
In this connection it is worth remarking that the possibility of a $s$-quark contribution produces considerable uncertainty in estimates of nucleon EDMs \cite{Chien:2015xha}, largely because its size is very uncertain \cite{Bhattacharya:2015esa}.  Therefore, even a rudimentary DSE estimate of this contribution could be useful.  Such may be obtained via a simplified treatment of meson-loop corrections to the quark gap equations, as used elsewhere \cite{Cloet:2008fw, Chang:2009ae} to estimate the proton's strangeness-magnetic-moment and -$\sigma$-term.  Following that reasoning, one is led to $\delta_T s(\zeta_2) \approx 0.02 g_T^{0} = 0.009$.

\section{Conclusion}
\label{SecConclusion}
We employed a confining, symmetry-preserving, Dyson-Schwinger equation treatment of a vector$\,\otimes\,$vector contact interaction in order to formulate and solve Faddeev equations for the nucleon and $\Delta$-baryon in which the kernel involves dynamical dressed-quark exchange.  These are the first contact-interaction calculations to produce momentum-dependent Faddeev amplitudes.  Previous contact-interaction studies have imposed a supplementary condition on the Faddeev equation kernel, \emph{viz}.\ a ``static approximation'' [Eq.\,\eqref{staticA}], which leads to momentum-independent amplitudes.

So far as computed masses of the nucleon and $\Delta$-baryon are concerned, eliminating the static approximation has little effect [Table~\ref{TabFEMasses}].   On the other hand, the impact on the Faddeev amplitudes is dramatic.  In stark contrast to static approximation results, contact-interaction dynamical quark exchange produces amplitudes that compare far more favourably with those obtained using kernels built from elements that possess momentum-dependence typical of QCD [Fig.\,\ref{F3}].  This marked improvement can potentially lead to a better description of those baryon properties which are sensitive to momentum scales that exceed the dressed-quark mass, $M\sim 0.4\,$GeV.  That can be tested, \emph{e.g}., in a study of the nucleon's first radial excitation and the related transition form factors, for which the contact-interaction$+$static-approximation yields poor results.

Regarding the proton's $\sigma$-term, the contact interaction result is one-half that obtained using more realistic interactions [Eq.\,\eqref{eqNsigmaF}], whether or not the static approximation is employed.  The small value is thus a defect of the contact interaction itself, which may be traced to the rigid nature of the diquark masses and Bethe-Salpeter amplitudes obtained with this interaction.

Implementing dynamical quark exchange in the contact-interaction Faddeev equation kernels yields results for the large Bjorken-$x$ values of the separate ratios of unpolarised and longitudinally-polarised valence $u$- and $d$-quark parton distribution functions that are noticeably different from those obtained using the static approximation.  The differences owe primarily to an increase in strength for the scalar diquark component of the nucleon.  Our dynamical results [Eqs.\,\eqref{largexB}] should be viewed as a truer representation of the contact-interaction's predictions.

The increase in strength for the scalar-diquark component of the nucleon, induced by dynamical quark exchange, also affects the computed values of the proton's tensor charges, $\delta_T u$, $\delta_T d$ [reported in Eq.\,\eqref{tensorz2} and compared with other estimates in Fig.\,\ref{TensorAll}]: it acts primarily to increase $\delta_T u$.  Our calculations confirm that the presence of diquark correlations in the proton reduces the size of $\delta_T u$, $\delta_T d$ compared with results obtained in simple quark models whilst simultaneously increasing $|\delta_T d/\delta_T u|$; verify that $\delta_T u$ is a direct measure of the strength of DCSB in the Standard Model; and emphasise that $|\delta_T d/\delta_T u|$ diminishes with $P^{p,a}/P^{p,s}$, \emph{i.e}.\ the ratio of pseudovector- and scalar-diquark interaction probabilities.

With this analysis we have completed the first improvement promised in Ref.\,\cite{Pitschmann:2014jxa} and so must address the question posed in the Introduction, \emph{viz}.\ Does the increased complexity which accompanies dynamical quark exchange in the Faddeev kernel outweigh the loss of simplicity inherent in using the static approximation?  If one is driving toward a realistic picture of a wide range of hadron physics observables, including those sensitive to probe momenta greater than the dressed-quark mass, then the answer must be affirmative.  This question has a natural extension, however: Is there merit in continuing to use a contact interaction when computational resources are beginning to enable the use of realistic interactions in the study of baryons?  Here the answer is:
%
Yes, depending upon the problem in hand; but the character of those problems is rapidly evolving.

Finally, returning to the proton's tensor charges, the next step should be computation using the approaches of Refs.\,\cite{Eichmann:2011vu, Segovia:2014aza, Segovia:2015ufa} in order to obtain continuum predictions with a direct connection to QCD.

\section*{Acknowledgments}
We are grateful for insightful comments and suggestions from M.~Pitschmann, S.-X.~Qin and S.-L.~Wan.
J.\,Segovia acknowledges financial support from a postdoctoral IUFFyM contract at the
Universidad de Salamanca.
Work also supported by:
the National Natural Science Foundation of China (grant nos.\ 11275097, 11475085 and 11535005);
the Fundamental Research Funds for the Central Universities Programme of China (grant no.\ WK2030040050);
and U.S.\ Department of Energy, Office of Science, Office of Nuclear Physics, under contract no.~DE-AC02-06CH11357.

\appendix
\setcounter{figure}{0}
\setcounter{table}{0}
\renewcommand{\thefigure}{\Alph{section}.\arabic{figure}}
\renewcommand{\thetable}{\Alph{section}.\arabic{table}}

\section{Collected Formulae}
\label{AppFormulae}
The matrices in Eqs.\,\eqref{SAStructure}, \eqref{Dstructure}, which express the Dirac-matrix structure of the positive-energy nucleon and $\Delta$, are
\begin{equation}
\label{tauN}
\begin{array}{ll}
\tau^1(l;P)= {\mathbf I}_{4\times 4}\,,
&\tau^2(l;P)=i\gamma\cdot{\hat{l}}^\perp \,, \\
\tau_\mu^3(l;P)=\frac{1}{\sqrt{3}}\gamma^\perp_\mu\gamma_5\,,
&\tau_\mu^4(l;P)=\frac{i}{\sqrt{3}}  \gamma^\perp_\mu \gamma\cdot{\hat{l}}^\perp\!\! \gamma_5\,,
\\
\tau_\mu^5(l;P)=-i \hat P_\mu\gamma_5\,,
&\tau_\mu^6(l;P)= \hat P_\mu \gamma\cdot{\hat{l}}^\perp \gamma_5\,,
\\
& \rule{-6em}{0ex} \tau_\mu^7(l;P)=\frac{1}{\sqrt{6}} (\gamma^\perp_\mu - 3 \gamma\cdot{\hat{l}}_T \hat l^\perp_\mu)\gamma_5\,,  \\
&\rule{-6em}{0ex} \tau_\mu^8(l;P)=\frac{i}{\sqrt{6}}  (\gamma^\perp_\mu \gamma\cdot{\hat{l}}_T - \hat l^\perp_\mu)\gamma_5 \,,
\end{array}
\end{equation}
where $\hat l^\perp_\nu = \hat l_\nu + \hat l\cdot\hat P\, \hat P_\nu$, $ \gamma^\perp_\nu = \gamma_\nu + \gamma\cdot\hat P\, \hat P_\nu$, $\hat l^2=1$, $\hat P^2= - 1$; and
{\allowdisplaybreaks
\begin{align}
\nonumber
\tau_{\nu\rho}^1(l;P)&= \delta_{\mu\nu}\,,\\
\nonumber
\tau_{\nu\rho}^2(l;P)&=\frac{i}{\sqrt{5}} \left( 2\gamma^\perp_\nu l_T^\rho -3 \delta_{\nu\rho} \gamma\cdot l^\perp \right),\\
\nonumber
\tau_{\nu\rho}^3(l;P)&=-i\sqrt{3}\hat P_\mu l^\perp_\nu \gamma\cdot{l}_T \,,\\ \tau_{\nu\rho}^4(l;P)&=\sqrt{3}\hat P_\mu \hat l^\perp_\nu\,,
\label{tauD}\\
\nonumber
\tau_{\nu\rho}^5(l;P) &= \gamma^\perp_\mu \hat l^\perp_\nu \gamma\cdot{\hat l}^\perp ,\\ \nonumber
\tau_{\nu\rho}^6(l;P) &= -i\gamma^\perp_\mu \hat l^\perp_\nu \,,\\
\nonumber
\tau_{\nu\rho}^7(l;P)&=-\gamma^\perp_\mu \hat l^\perp_\nu \gamma\cdot{\hat l}^\perp -\delta^{\mu\nu} + 3\hat l^\perp_\mu \hat l^\perp_\nu \,, \\
\nonumber
\tau_{\nu\rho}^8(l;P)&=\frac{i}{\sqrt{5}}\left( \delta_{\mu\nu} \gamma\cdot{\hat l}^\perp + \gamma^\perp_\mu \hat l^\perp_\nu -5\hat l^\perp_\mu \hat l^\perp_\nu \gamma\cdot{\hat l}^\perp \right).
\end{align}
}

We using the following projection operators in arriving at the scalar-valued integral equations in Sec.\,\ref{DefineFE} -- for the nucleon,
\begin{equation}
\label{ProjectionN}
\rule{-0.7em}{0ex}\begin{array}{ll}
\bar\tau^1(l;P)=\frac{1}{2} {\mathbf I}_{4\times 4}\,,
&\bar\tau^2(l;P)=-\frac{i}{2}\gamma\cdot\hat{l}^\perp ,\\
\bar\tau_\mu^3(l;P)=\frac{1}{2\sqrt{3}}\gamma_5 \gamma^\perp_\mu\,,
&\bar\tau_\mu^4(l;P)=-\frac{i}{2\sqrt{3}}\gamma_5 \gamma^\perp_\mu \gamma\cdot\hat{l}^\perp,
\\
\bar\tau_\mu^5(l;P)=-\frac{1}{2}i\gamma_5\hat P_\mu\,,
&\bar\tau_\mu^6(l;P)=-\frac{1}{2}\gamma_5\hat P_\mu \gamma\cdot\hat{l}^\perp,
\\
& \rule{-6em}{0ex} \bar\tau_\mu^7(l;P)=\frac{1}{2\sqrt{6}}\gamma_5(\gamma^\perp_\mu - 3 \gamma\cdot\hat{l}^T \hat l^T_\mu),\\
& \rule{-6em}{0ex}  \bar\tau_\mu^8(l;P)=-\frac{i}{2\sqrt{6}}\gamma_5 (\gamma^T_\mu \gamma\cdot\hat{l}_T - \hat l^\perp_\mu);
\end{array}
\end{equation}
and for the $\Delta$-baryon,
{\allowdisplaybreaks
\begin{align}
\nonumber
&\bar\tau_{\nu\rho}^1(l;P)= \frac{1}{4}\delta_{\mu\nu} \,,\\
\nonumber
&\bar\tau_{\nu\rho}^2(l;P)=-\frac{i}{4\sqrt{5}}\left( 2\gamma^\perp_\nu l^\perp_\rho -3 \delta_{\nu\rho} \gamma\cdot l^\perp \right)\,, \\
\nonumber
&\bar\tau_{\nu\rho}^3(l;P)=-i\frac{\sqrt{3}}{4}\hat P_\mu l^\perp_\nu \gamma\cdot{l}^\perp,\\
&\bar\tau_{\nu\rho}^4(l;P)=-\frac{\sqrt{3}}{4}\hat P_\mu \hat l^\perp_\nu\,,
\label{ProjectionD}\\
\nonumber
& \bar\tau_{\nu\rho}^5(l;P)= \frac{1}{4}\hat l^\perp_\nu  \gamma\cdot\hat{l}^\perp\gamma^\perp_\mu\,,\\
\nonumber
&\bar\tau_{\nu\rho}^6(l;P)=\frac{i}{4}\gamma^\perp_\mu \hat l^\perp_\nu\,,\\
\nonumber
&\bar\tau_{\nu\rho}^7(l;P)=-\frac{1}{4} \hat l^\perp_\nu \gamma\cdot\hat {l}^\perp\gamma^\perp_\mu -\delta^{\mu\nu} + 3\hat l^\perp_\mu\hat l^\perp_\nu\,.\\
\nonumber
&\bar\tau_{\nu\rho}^8(l;P)=-\frac{i}{4\sqrt{5}}\left( \delta_{\mu\nu}  \gamma\cdot\hat{l}^\perp + \gamma^\perp_\mu \hat l^\perp_\nu -5\hat l^\perp_\mu\hat l^\perp_\nu \gamma\cdot\hat {l}^\perp \right)
\end{align}}

These projectors are defined such that, for the nucleon:
{\allowdisplaybreaks
\begin{subequations}
\begin{align}
&{\rm tr}\left[\bar\tau^i(l)\tau^j(l)\Lambda_+(P)\right]=\delta^{ij},\;i,j=1,2\,,
\\
&{\rm tr}\left[\bar\tau_\mu^i(l)\tau_\mu^j(l)\Lambda_+(P)\right]=\delta^{ij},\;
i,j=3,\ldots,8\,,
\\
&{\rm tr}\left[\bar\tau^i(l)\tau_\mu^j(l)\Lambda_+(P)\right]=0\,,\;i=1,2, j=3,\ldots,8\,,
\end{align}
\end{subequations}}
\hspace*{-0.5\parindent}where the positive-energy projector is defined via
\begin{equation}
\label{Lplus} 2 M \, \Lambda_+(P):= \sum_{s=\pm} \, u(P,s) \, \bar
u(P,s) = \left( -i \gamma\cdot P + M\right).
\end{equation}

In connection with the $\Delta$-baryon,
\begin{eqnarray}
Tr\left[\bar\tau^i_{\mu\nu}(l)\tau^j_{\mu\rho}(l) \mathcal{R}^{\Delta}_{\rho\nu}(P)\right]=\delta^{ij}, i,j=1,\ldots,8\,,
\end{eqnarray}
where
\begin{align}
\nonumber
\mathcal{R}&^\Delta_{\mu\nu}(P)=[\delta_{\mu\nu} \\
&-\tfrac{1}{3}\gamma_\mu\gamma_\nu+\tfrac{2}{3}\hat{P}_\mu\hat{P}_\nu - \tfrac{i}{3}(\hat{P}_\mu\gamma_\nu-\hat{P}_\nu\gamma_\mu)]\Lambda_+(P).
\end{align}

\section{Nucleon-Photon Vertex}
\label{NPVertex}
\subsection{Diagram 1: Normalisation}
\label{DiagramOne}
When a symmetry-preserving regularisation of the contact interaction is employed, one finds
\begin{equation}
\Gamma_\mu^\gamma(Q^2=0) = \gamma_\mu\,,
\end{equation}
\emph{i.e}.\ the dressed-photon--quark vertex preserves its bare form at zero momentum-transfer \cite{Roberts:2010rn}.  Consequently, the ${\mathpzc C}^{1}_{00}$ (scalar-diquark) contribution to the nucleon normalisation constant, ${\mathpzc N}_p$, can be written
\begin{align}
\nonumber
& Q_{00}^1 \Lambda_+(P)\gamma_\mu \Lambda_+(P)\!\!
={\mathpzc N}_p \Lambda_+(P) \int_{dk}^\Lambda \overline{\mathcal S}(k;-P) \\
&
\times S(k_q) e_u \gamma_\mu S(k_q) \Delta^{0^+}(k_{qq}) {\mathcal S}(k;P) \Lambda_+(P)\,,
\label{Diagram100}
\end{align}
where ${\mathcal S}(k;P)$ is the $[ud]$ scalar-diquark component of the nucleon's Faddeev amplitude in Eq.\,\eqref{SStructure} and $\overline{\mathcal S}(k;-P) =C^\dagger\overline{\mathcal S}(-k;-P)C$; $S(k_q)$ is the dressed-quark propagator described in Sec.\,\ref{SecGap}; and $e_u= (2/3)$.

After some algebra, following the pattern in Sec.\,\ref{SecRegEq}, Eq.\,\eqref{Diagram100} yields
\begin{equation}
\label{Q001}
Q_{00}^1  = 2 e_u {\mathpzc N}_p
\int_0^1 dx\,x\int_{dk}^\infty{\mathpzc E}_2(k^2+\omega_0)
{\mathpzc D}_{00}^1(\tilde k;P)\,,
\end{equation}
where, with ${\mathpzc n}(\ell) = -i\gamma\cdot\ell + M$,
\begin{subequations}
\begin{align}
\nonumber
&{\mathpzc D}_{00}^1(k;P) =
{\rm tr}  \tfrac{1}{2}\gamma_\mu \Lambda_+(P)
\\
&\quad\quad\quad \times \overline{\mathcal S}(k;-P){\mathpzc n}(k_q) \gamma_\mu{\mathpzc n}(k_q) {\mathcal S}(k;P) \Lambda_+(P)\,,\\
&{\tilde{k}} =k +(\tfrac{2}{3}-x)P \,, \\
&\omega_0  =xM^2+(1-x)m_{qq0^+}^2-x(1-x)m_N^2 \,.
\end{align}
\end{subequations}

Inserting our $g_N=1.28$ computed nucleon mass and Faddeev amplitude into Eq.\,\eqref{Q001}, we obtain
\begin{equation}
\label{D1}
Q_{00}^1 = 0.0163 \, e_u {\mathpzc N}_p =: {\mathpzc Q}^0 e_u {\mathpzc N}_p\,.
\end{equation}

Diagram~1 also represents two other cases, \emph{viz}.\ a dressed-quark struck with either $\{ud\}$ or $\{uu\}$ axial-vector diquarks as the spectator, ${\mathpzc C}^1_{\rightarrow\rightarrow}$, ${\mathpzc C}^1_{\uparrow\uparrow}$ respectively.  Consider first the case of a $u$-quark struck with a $\{ud\}$ spectator.  Analysis of the type described above yields
\begin{align}
\label{Q111}
Q_{11}^1 & = e_u {\mathpzc N}_p
\int_0^1 dx\,x\int_{dk}^\infty{\mathpzc E}_2(k^2+\omega_1)
{\mathpzc D}_{11}^1(\tilde k;P)\,,
\end{align}
where
\begin{subequations}
\begin{align}
\nonumber
&{\mathpzc D}_{11}^1(k;P) =
{\rm tr}  \tfrac{1}{2}\gamma_\mu \Lambda_+(P)\overline{\mathcal A}_\alpha(k;-P) \\
&\quad \times {\mathpzc n}(k_q) \gamma_\mu {\mathpzc n}(k_q) T^1_{\alpha\beta}(k_{qq})
{\mathcal A}_\beta(k;P) \Lambda_+(P)\,,\\
&\omega_1  =xM^2+(1-x)m_{qq1^+}^2-x(1-x)m_N^2 \,.
\end{align}
\end{subequations}

Inserting our $g_N=1.28$ nucleon mass and Faddeev amplitude into Eq.\,\eqref{Q111}, we obtain
\begin{equation}
\label{D2}
Q_{11}^1 = 0.00148 \, e_u {\mathpzc N}_p =: {\mathpzc Q}^1 e_u {\mathpzc N}_p\,.
\end{equation}

Owing to isospin symmetry, the other case, $d$-quark struck with a $\{uu\}$ spectator, ${\mathpzc C}^1_{\uparrow\uparrow}$, contributes $(e_d = -e_u/2)$
\begin{equation}
\label{D3}
(e_d/e_u) Q_{11}^1 (-\sqrt 2)^2 = -Q_{11}^1.
\end{equation}
Plainly, therefore, diagrams with axial-vector diquark spectators do not contribute to the proton's normalisation.  More generally, in fact, they contribute nothing to the proton's electromagnetic form factors \cite{Wilson:2011aa}: ${\mathpzc C}^1_{\rightarrow\rightarrow}+{\mathpzc C}^1_{\uparrow\uparrow}\equiv 0$.

\subsection{Diagram 2: Normalisation}
\label{DiagramTwo}
In this instance, the first contribution we consider is a $u$-quark spectator to a photon--scalar-diquark interaction, ${\mathpzc C}^2_{00}$:
\begin{align}
\nonumber
& \rule{-0.5em}{0ex} Q^2_{00} \Lambda_+(P)\gamma_\mu \Lambda_+(P)\!\!
={\mathpzc N}_p \Lambda_+(P)
\int_{dk}^\Lambda \overline{\mathcal S}(k;-P)\\
& \rule{-0.5em}{0ex}\times \Delta^{0^+}(k_{qq}) \mathcal{V}_\mu(k_{qq}) \Delta^{0^+}(k_{qq}) S(k_q) {\mathcal S}(k;P) \Lambda_+(P)\,,
\label{D200}
\end{align}
where, owing to a Ward-Green-Takahashi identity maintained by a symmetry-preserving regularisation of the contact interaction \cite{Roberts:2011wy}, the dressed photon--scalar-diquark vertex at zero momentum transfer is $(e_{ud} = e_u+e_d)$
\begin{equation}
\mathcal{V}_\mu(k_{qq}) = 2 \, e_{ud} \, k_{qq\,\mu}\,.
\end{equation}
Using this result and the now standard procedures, Eq.\,\eqref{D200} yields
\begin{equation}
\label{D4}
Q^2_{00} = 0.0152 \,e_{ud}\,{\mathpzc N}_p =: {\mathpzc D}^0 e_{ud}\,{\mathpzc N}_p\,.
\end{equation}

Diagram~2 also represents two other cases, \emph{viz}.\ a dressed-quark spectator to a photon interacting with either a $\{ud\}$ or $\{uu\}$ axial-vector diquark, ${\mathpzc C}^2_{\rightarrow\rightarrow}$, ${\mathpzc C}^2_{\uparrow\uparrow}$ respectively.  The $u$-quark spectator contribution is:
\begin{align}
\nonumber
& \rule{-0.5em}{0ex} Q^2_{11} \Lambda_+(P)\gamma_\mu \Lambda_+(P)\!\!
={\mathpzc N}_p \Lambda_+(P)
\int_{dk}^\Lambda \overline{\mathcal A}_\alpha(k;-P)\Delta^{1^+}_{\alpha\alpha^\prime}(k_{qq}) \\
& \rule{-0.5em}{0ex}\times \mathcal{V}_{\mu,\alpha^\prime\beta^\prime}(k_{qq}) \Delta^{1^+}_{\beta^\prime\beta}(k_{qq}) S(k_q) {\mathcal A}(k;P) \Lambda_+(P)\,,
\label{D211}
\end{align}
which involves the dressed photon--pseudovector-diquark vertex at zero momentum transfer \cite{Roberts:2011wy}:
\begin{equation}
\mathcal{V}_{\mu,\alpha\beta}(k_{qq}) =
2 \, e_{ud} k_{qq\,\mu} T^1_{\alpha\beta}(k_{qq})\,.
\end{equation}
Using this result, Eq.\,\eqref{D211} yields
\begin{equation}
\label{D5}
Q^2_{11} = 0.000879 \,e_{ud}\,{\mathpzc N}_p =: {\mathpzc D}^1 e_{ud}\,{\mathpzc N}_p\,.
\end{equation}
Owing to isospin symmetry, the analogous result for ${\mathpzc C}^2_{\uparrow\uparrow}$, the $d$-quark spectator diagram, is $(e_{uu}=2 e_u=4e_{ud})$
\begin{equation}
\label{D6}
(e_{uu}/e_{ud}) (-\sqrt 2)^2 Q^2_{11} = 8 {\mathpzc D}^1 e_{ud}\,{\mathpzc N}_p\,.
\end{equation}

\subsection{Diagram 3: Normalisation}
\label{DiagramThree}
If quark flavours are distinguished, then this image in Fig.\,\ref{CIcurrent} corresponds to eight two-loop diagrams.  The simplest example involves a scalar-diquark in the initial and final states, so that a dressed $d$-quark is struck ``in-flight'' by the photon, ${\mathpzc C}^3_{00}$:
\begin{align}
\nonumber
& Q^3_{00} \Lambda_+(P)\gamma_\mu \Lambda_+(P) =
{\mathpzc N}_p \Lambda_+(P) e_d g_N^2 \!\! \int_{dk}^\Lambda \int_{dl}^\Lambda
\overline{\mathcal S}(k;-P) \\
\nonumber
& \times  S(k_q) \Delta^{0^+}(k_{qq}) \Gamma^{0^+}(l_{qq}) [S(-k_X) \gamma_\mu S(-k_X)]^{\rm T} \\
& \times \bar \Gamma^{0^+}(-k_{qq}) S(l_q) \Delta^{0^+}(l_{qq}){\mathcal S}(l;P) \Lambda_+(P)\,,
\label{Q300}
\end{align}
where $k_X = k + l - P/3$.  After some algebra, Eq.\,\eqref{Q300} can be recast in the following form:
\begin{equation}
Q^3_{00} =  e_d g_N^2 {\mathpzc N}_p \!\! \int_{dk}^\Lambda \int_{dl}^\Lambda
\frac{{\mathpzc N}^3_{\;00}}{D_1 D_2 D_3 D_4 D_5^2}\,,
\label{Q300A}
\end{equation}
with
\begin{align}
\nonumber
D_1 & = k_q^2 + M^2, & D_2 = k_{qq}^2 + m_{qq0^+}^2,\\
D_3 & = l_q^2 + M^2, & D_4 = l_{qq}^2 + m_{qq0^+}^2,\\
\nonumber
D_5 & = k_X^2 + M^2,
\end{align}
and
\begin{align}
\nonumber
{\mathpzc N}^3_{\;00} & = {\rm tr}  \tfrac{1}{2}\gamma_\mu \Lambda_+(P) \overline{\mathcal S}(k;-P)
S(k_q) \Delta^{0^+}(k_{qq}) \Gamma^{0^+}(l_{qq}) \\
\nonumber
& \times  [{\mathpzc n}(-k_X) \gamma_\mu {\mathpzc n}(-k_X)]^{\rm T}  \bar \Gamma^{0^+}(-k_{qq}) \\
& \times {\mathpzc n}(l_q) \Delta^{0^+}(l_{qq}){\mathcal S}(l;P) \Lambda_+(P)\,.
\end{align}
In the algebraic evaluation of ${\mathpzc N}^3_{\;00}$, the following replacements are used sequentially:
\begin{equation}
\label{onshellreplacements}
\begin{array}{ll}
k_{qq}^2 \to -m_{qq0^+}^2,\; & l_{qq}^2 \to -m_{qq0^+}^2,\; \\
k_{qq\,\mu} \to \tfrac{2}{3} P_\mu,\; & l_{qq\,\mu} \to \tfrac{2}{3} P_\mu.
\end{array}
\end{equation}

As usual, the next step is to combine the denominators using a Feynman parametrisation, in which case Eq.\,\eqref{Q300A} becomes
\begin{align}
\nonumber
& Q^3_{00}  = e_d g_N^2 {\mathpzc N}_p \!\!\int_{dk}^\Lambda \int_{dl}^\Lambda \\
& \times
24 \int_0^1 dx_1 dx_2 dx_3 dx_4 x_1^4 x_2^3 x_3^2 x_4 \frac{{\mathpzc N}^3_{\;00}(\tilde k,\tilde l;P)}{[D^3_{00}]^6}\,,
\end{align}
where the variable transformations $k\to \tilde k$, $l \to \tilde l$ are constructed so that $D^3_{00}={\mathpzc f}(k^2,l^2,P^2=-m_N^2)$.  At this point one can implement the confining regularisation prescription and thereby obtain
\begin{align}
\nonumber
& Q^3_{00}  = e_d g_N^2 {\mathpzc N}_p \!\!\int_{dk}^\infty \int_{dl}^\infty
24 \int_0^1 dx_1 dx_2 dx_3 dx_4 x_1^4 x_2^3 x_3^2 x_4 \, \\
& \times
{\mathpzc N}^3_{\;00}(\tilde k,\tilde l;P)\,{\mathpzc E}_5({\mathpzc f}(k^2,l^2,-m_N^2)\,. \label{Q300B}
\end{align}

The precise forms for the elements in Eq.\,\eqref{Q300B} are lengthy so we do not present them here.  Notwithstanding that, their computation is straightforward; and inserting our $g_N=1.28$ nucleon mass and Faddeev amplitudes into the expressions one derives, the following numerical result is obtained:
\begin{align}
Q^3_{00} = 5.34 \times 10^{-5} e_d g_N^2 {\mathpzc N}_p & = 8.75 \times 10^{-5} e_d {\mathpzc N}_p\\
& =: {\mathpzc X}^{00} e_d {\mathpzc N}_p\,.
\label{D7}
\end{align}

There are seven more contributions; but, owing to isospin symmetry, as described in connection with Eq.\,\eqref{CNisospin}, only two additional computations are necessary: $\{ud\}$ pseudovector-diquark breakup, $d$-quark struck in-flight, $[ud]$ scalar diquark recombination $([ud]\,d_\gamma\{ud\})$, ${\mathpzc C}^{3}_{0\rightarrow}$; and $\{ud\}$ pseudovector-diquark breakup, $d$-quark struck in-flight, $\{ud\}$ pseudovector diquark recombination $(\{ud\}\,d_\gamma\{ud\})$, ${\mathpzc C}^{3}_{\rightarrow\rightarrow}$.  The results are:
\begin{subequations}
\begin{align}
\label{D8}
 Q^3_{0\rightarrow} & = 1.79 \times 10^{-5} e_d {\mathpzc N}_p =: {\mathpzc X}^{0\rightarrow} e_d {\mathpzc N}_p\,,\\
\label{D9}
[ud]\,u_\gamma \{uu\} & = {\mathpzc C}^{3}_{0\uparrow }
 = -\sqrt{2} e_u {\mathpzc X}^{0\rightarrow} {\mathpzc N}_p\,, \\
\label{D10}
\{ud\}\,d_\gamma [ud] & = {\mathpzc C}^{3}_{\rightarrow 0} = e_d {\mathpzc X}^{0\rightarrow} {\mathpzc N}_p\,,\\
\label{D13}
\{uu\}\,u_\gamma [ud] & = {\mathpzc C}^{3}_{\uparrow 0}= -\sqrt{2} e_u {\mathpzc X}^{0\rightarrow} {\mathpzc N}_p\,;
\end{align}
\end{subequations}
and
\begin{subequations}
\begin{align}
\label{D11}
Q^3_{\rightarrow\rightarrow} & = -8.8 \times 10^{-7} e_d {\mathpzc N}_p =: {\mathpzc X}^{\rightarrow\rightarrow} e_d {\mathpzc N}_p \,,\\
\label{D12}
\{ud\}\,u_\gamma &\{uu\} = {\mathpzc C}^{3}_{\rightarrow\uparrow} = -\sqrt{2} e_u {\mathpzc X}^{\rightarrow\rightarrow}  {\mathpzc N}_p \,,\\
\label{D14}
\{uu\}\,u_\gamma &\{ud\} = {\mathpzc C}^{3}_{\uparrow\rightarrow} = -\sqrt{2} e_u {\mathpzc X}^{\rightarrow\rightarrow}  {\mathpzc N}_p \,.
\end{align}
\end{subequations}

\subsection{Collected Results: Normalisation}
For ease of reference, we gather all independent results computed in Secs.\,\ref{DiagramOne}--\ref{DiagramThree} into Eq.\,\eqref{collectedNorm}, wherein each entry should be divided by $10^{3}$:
\begin{equation}
\begin{array}{ccccccc}
{\mathpzc Q}^0 & {\mathpzc Q}^1 & {\mathpzc D}^0 & {\mathpzc D}^1 &
{\mathpzc X}^{00} &  {\mathpzc X}^{0\rightarrow} &  {\mathpzc X}^{00} \\
16.3 & 1.48 & 15.2 & 0.879 & 0.0875 & 0.0179 & -8.80 \times 10^{-4}
\end{array}\,.
\label{collectedNorm}
\end{equation}
The associated contributions to the nucleon's canonical normalisation constant are listed in Table~\ref{NormalisationContributions}.

\begin{table}[t]
\caption{\label{NormalisationContributions}
Breakdown by diagram of contributions to the nucleon's canonical normalisation.  Column 2: Results scaled as described in Sec.\,\ref{SecCC}.   \emph{N.B}.\ In relation to Eqs.\,\eqref{collectedNorm}, \eqref{protoncharge}: Diagram~1 corresponds to summing ${\mathpzc Q}$ terms; Diagram~2, summation of ${\mathpzc D}$ terms; and Diagram~3, summation of ${\mathpzc X}$ terms. (Actual value of each entry is obtained via division by $10^{3}$.)
}
\begin{center}
\begin{tabular*}
{\hsize}
{
l@{\extracolsep{0ptplus1fil}}
c@{\extracolsep{0ptplus1fil}}
c@{\extracolsep{0ptplus1fil}}}\hline
          & $1/{\mathpzc N}_{\;p}$ & $1/{\mathpzc N}_{\;p}^{\rm R}$ \\\hline
Diagram~1 & $10.9\phantom{00}$ & $10.5\phantom{000}$ \\
Diagram~2 & $\phantom{1}7.70\phantom{0}$ & $8.46\phantom{0}$ \\
Diagram~3 & $-0.0729$ & $0.157$ \\
Diagram~4 & $0\phantom{.00}$ & $0\phantom{.000}$ \\\hline
Total     & $18.5\phantom{00}$ & $19.2\phantom{00}$ \\\hline
\end{tabular*}
\end{center}
\end{table}

\subsection{Current Conservation}
\label{SecCC}
In a symmetry preserving treatment of the contact interaction, using integration by parts and changes of variables, one can establish the following Ward identities between the contributions described above:
\begin{subequations}
\label{NewWIs}
\begin{align}
\label{WI1}
{\mathpzc Q}^0 & = {\mathpzc D}^0 + 2 {\mathpzc X}^{00} + 2 (1-\sqrt{2}) {\mathpzc X}^{0\rightarrow}\,,\\
\label{WI2}
{\mathpzc Q}^1 & = {\mathpzc D}^1 + 2 {\mathpzc X}^{0\rightarrow} + 2 (1-\sqrt{2}) {\mathpzc X}^{\rightarrow\rightarrow}\,,\\
\label{WI3}
{\mathpzc X}^{\rightarrow\rightarrow} & = \frac{1+\sqrt{2}}{1-\sqrt{2}} \, {\mathpzc X}^{0\rightarrow}\,.
\end{align}
\end{subequations}
In the static approximation, Eqs.\,\eqref{staticA}, \eqref{staticAR}, the exchange diagrams vanish and Eqs.\,\eqref{NewWIs} reproduce the identities used in Refs.\,\cite{Wilson:2011aa, Segovia:2013rca, Segovia:2013uga}.

At the same level, the neutron should be neutral.  Consider, therefore, that in the isospin symmetric limit, the neutron's charge can be computed from the diagrams detailed in Secs.\,\ref{DiagramOne}--\ref{DiagramThree} by making the exchange $u\leftrightarrow d$:
{\allowdisplaybreaks
\begin{align}
\nonumber
 e_n & = {\mathpzc N}_p [
{\mathpzc Q}^0 e_d + {\mathpzc Q}^1 (e_d+2 e_u) \\
\nonumber
& \quad +
{\mathpzc D}^0 e_{ud} + {\mathpzc D}^1 (e_{ud} + 2 e_{dd})\\
\nonumber & \quad +
{\mathpzc X}^{00} e_u + 2 {\mathpzc X}^{0\rightarrow} (e_u - \sqrt{2} e_d)\\
& \quad
+ {\mathpzc X}^{\rightarrow\rightarrow}(e_u - 2 \sqrt{2} e_d)
] =0 \,,
\end{align}}
\hspace*{-0.5\parindent}if Eqs.\,\eqref{NewWIs} are satisfied.  Numerically, however, they are not.  That is because the last step in our regularisation procedure involves the introduction of hard infrared and ultraviolet cutoffs, which naturally spoils the conditions necessary to prove Eqs.\,\eqref{NewWIs}.  Following Ref.\,\cite{Pitschmann:2014jxa}, a remedy is straightforward.  Namely, we use Eq.\,\eqref{WI3} as a definition of ${\mathpzc X}^{11}$ and introduce rescaling factors ${\mathpzc r}_{0,1}$, with
${\mathpzc Q}^0 \to \bar {\mathpzc Q}^0={\mathpzc Q}^0 (1+ {\mathpzc r}_0)$,
${\mathpzc D}^0 \to \bar {\mathpzc D}^0 = {\mathpzc D}^0 (1- {\mathpzc r}_0)$
and
${\mathpzc Q}^1 \to \bar{\mathpzc Q}^1 ={\mathpzc Q}^1 (1+ {\mathpzc r}_1)$,
${\mathpzc D}^1 \to {\mathpzc D}^1={\mathpzc D}^1 (1- {\mathpzc r}_1)$,
such that Eqs.\,\eqref{NewWIs} are satisfied when expressed in terms of $\bar{\mathpzc Q}^{0,1}$, $\bar{\mathpzc D}^{0,1}$.  This procedure yields
\begin{equation}
{\mathpzc r}_{0} = -0.0298\,,\;
{\mathpzc r}_{1} = -0.203\,,\;
\end{equation}
and entails ${\mathpzc X}^{\rightarrow\rightarrow} \to \bar {\mathpzc X}^{\rightarrow\rightarrow} = 119 {\mathpzc X}^{0\rightarrow}=:{\mathpzc r}_{\mathpzc X}{\mathpzc X}^{0\rightarrow}$. \emph{N.B}.\ These same scaling factors are applied to the analogous diagrams that appear in computing the proton's tensor charges.

The nucleon Faddeev amplitude can now be normalised, following the procedure described in Sec.\,\ref{SecCN}.  One has
{\allowdisplaybreaks
\begin{align}
\nonumber
 e_p & = {\mathpzc N}_p [
\bar {\mathpzc Q}^0 e_u + \bar {\mathpzc Q}^1 (e_u+2 e_d) \\
\nonumber
& \quad +
\bar {\mathpzc D}^0 e_{ud} + \bar {\mathpzc D}^1 (e_{ud} + 2 e_{uu})\\
\nonumber & \quad +
{\mathpzc X}^{00} e_d + 2 {\mathpzc X}^{0\rightarrow} (e_d - \sqrt{2} e_u)\\
& \quad
+ \bar {\mathpzc X}^{\rightarrow\rightarrow}(e_d - 2 \sqrt{2} e_u)
] = 0.0192\,{\mathpzc N}_p
\label{protoncharge}\\
& \Rightarrow {\mathpzc N}_p = 52.2\,.
\end{align}}
\hspace*{-0.5\parindent}The contribution from each diagram-number is listed separately in Table~\ref{NormalisationContributions}.

\section{Nucleon-Tensor Vertex}
\label{PTensor}
When a symmetry-preserving regularisation of the contact interaction is employed, one finds \cite{Pitschmann:2014jxa}
\begin{equation}
{\mathpzc V}_{\mu\nu} = \sigma_{\mu\nu}\,,
\end{equation}
\emph{i.e}.\ the tensor vertex is not dressed, a result which owes to the inability of any symmetry-preserving regularisation of a contact interaction to support nonzero relative momentum in a quark-antiquark system.

Following our analysis of the proton's electromagnetic current at $Q^2=0$, only one type of diagram remains to be explicated, \emph{viz}. Diagram 4 in Fig.\,\ref{CIcurrent}.  In connection with a zero-momentum tensor probe, this image translates into two terms: ${\mathpzc C}^4_{0\rightarrow}={\mathpzc C}^4_{\rightarrow 0}$, with the latter corresponding to
\begin{align}
\nonumber
&{\mathpzc T}^4_{0\rightarrow} \Lambda_+(P)\sigma_{\mu\nu}\Lambda_+(P)
= {\mathpzc N}_p \Lambda_+(P) \int_{dk}^\Lambda
\overline{\mathcal A}_\alpha(k;-P)  \\
\nonumber
& \times \Delta^{1^+}_{\alpha\beta}(k_{qq1^+})\mathcal{V}_{\beta\mu\nu}(k_{qq1^+},k_{qq0^+}) \Delta^{0^+}(k_{qq0^+}) \\
& \times S(k_q){\mathcal S}(k;P)\Lambda_+(P)\,,
\end{align}
where the tensor transition vertex is ($N_f = 2$, $N_c^D=2$):
\begin{align}
\nonumber
\mathcal{V}_{\beta\mu\nu}&(k_{qq1^+},k_{qq0^+})=-2 N_f N_c^D
{\rm tr}\int_{dq}^\Lambda \,  S(q+k_{qq1^+})\sigma_{\mu\nu} \\
& \times S(q+k_{qq0^+}) \Gamma^{0^+}(k_{qq0^+}) S(q) \overline{\Gamma}^{1^+}_\beta(-k_{qq1^+}) \,.
\end{align}

Analysing and combining these integrals in the now customary way, then at the hadronic scale $\zeta_H=0.39\pm0.02\,$GeV one obtains
\begin{equation}
{\mathpzc T}^4_{0\rightarrow} = -3.68 \times 10^{-3} {\mathpzc N}_p = -0.192\,.
\end{equation}
Both ${\mathpzc C}^4_{0\rightarrow}$, ${\mathpzc C}^4_{\rightarrow 0}$ contribute half their strength equally to $\delta_T u$, $\delta_T d$, so that
\begin{equation}
\label{Tensor4All}
\delta_T^{4} u = \delta_T^{4} d = {\mathpzc T}^4_{0\rightarrow}\,.
\end{equation}

The complete list of contributions from Diagrams~1-3 is:
{\allowdisplaybreaks
\begin{subequations}
\label{Tensor1All}
\begin{align}
{\mathpzc T}^1_{00} & = 0.0147 \, {\mathpzc N}_p \, (1+{\mathpzc r}_0) = 0.744 \,,\\
{\mathpzc T}^1_{\rightarrow\rightarrow} & = -4.86 \times 10^{-4} \, {\mathpzc N}_p \, (1+{\mathpzc r}_1) = -0.0195\,,\\
{\mathpzc T}^1_{\uparrow\uparrow} & = 2 {\mathpzc T}^1_{\rightarrow\rightarrow}\,, 
\end{align}
\end{subequations}
\begin{subequations}
\label{Tensor2All}
\begin{align}
{\mathpzc T}^2_{00} & = 0 \,,\\
{\mathpzc T}^2_{\rightarrow\rightarrow} & = 8.48\times 10^{-4} \, {\mathpzc N}_p \, (1-{\mathpzc r}_1) = 0.0544\,, \\
{\mathpzc T}^2_{\uparrow\uparrow} & = 2 {\mathpzc T}^2_{\rightarrow\rightarrow}\,,
\end{align}
\end{subequations}
\begin{subequations}
\label{Tensor3All}
\begin{align}
{\mathpzc T}^3_{00} & = -1.93 \times 10^{-4}\,{\mathpzc N}_p  = -0.00992\,,\\
{\mathpzc T}^3_{0\rightarrow} & = 8.8 \times 10^{-5} \,{\mathpzc N}_p = 0.00459\,,\\
{\mathpzc T}^3_{0\uparrow} & = -\sqrt{2} {\mathpzc T}^3_{0\rightarrow}\,,\\
{\mathpzc T}^3_{\rightarrow 0} & = {\mathpzc T}^3_{0\rightarrow}\,,\\
{\mathpzc T}^3_{\uparrow 0 } & = -\sqrt{2}{\mathpzc T}^3_{0\rightarrow}\,,\\
{\mathpzc T}^3_{\rightarrow \rightarrow} & = -7.87 \times 10^{-6}\,{\mathpzc N}_p \, {\mathpzc r}_{\mathpzc X}  = -0.0487\,, \\
{\mathpzc T}^3_{\rightarrow \uparrow} & = -\sqrt{2} {\mathpzc T}^3_{\rightarrow \rightarrow} \,,\\
{\mathpzc T}^3_{\uparrow \rightarrow } & = -\sqrt{2}{\mathpzc T}^3_{\rightarrow \rightarrow}\,.
\end{align}
\end{subequations}}
\hspace*{-0.5\parindent}In terms of the contributions in Eqs.\,\eqref{Tensor4All}--\eqref{Tensor3All}, the proton's tensor charges are:
\begin{subequations}
\begin{align}
\nonumber
\delta_T u & = [{\mathpzc T}^1_{00} + {\mathpzc T}^1_{\rightarrow\rightarrow}]
+ [\tfrac{1}{2}{\mathpzc T}^2_{\rightarrow\rightarrow} + {\mathpzc T}^2_{\uparrow\uparrow}] \\
& \quad + [{\mathpzc T}^3_{0\uparrow}+{\mathpzc T}^3_{\rightarrow\uparrow} + {\mathpzc T}^3_{\uparrow 0} + {\mathpzc T}^3_{\uparrow\rightarrow}] + {\mathpzc T}^4_{\rightarrow 0}\,,\\
\nonumber
\delta_T d & = {\mathpzc T}^1_{\uparrow\uparrow} + \tfrac{1}{2}{\mathpzc T}^2_{\rightarrow\rightarrow} \\
& \quad + [{\mathpzc T}^3_{00} + {\mathpzc T}^3_{0\rightarrow} + {\mathpzc T}^3_{\rightarrow 0} + {\mathpzc T}^3_{\rightarrow\rightarrow}] + {\mathpzc T}^4_{\rightarrow 0}\,.
\end{align}
\end{subequations}
These expressions yield the results in the upper panel of Table~\ref{TensorContributions}.


\end{document}